\documentclass[12pt,letterpaper]{article}
\usepackage[includeheadfoot,  
            marginratio={1:1,2:3},  
            width=412pt,  
            height=688pt,]{geometry} 
\usepackage{annulus} 
\usepackage{amsmath} 
\usepackage{amsfonts} 
\usepackage{amssymb} 
\usepackage{bbm} 
 
\usepackage{graphicx} 
\usepackage{subfigure}

\allowdisplaybreaks[2] 
\numberwithin{equation}{section} 
 
\newcommand{\thbb}[2]{\vartheta\genfrac[]{0pt}{}{#1}{#2}}
 
\newcommand{\be}{\begin{equation}} 
\newcommand{\ee}{\end{equation}} 
\newcommand{\bea}{\begin{eqnarray}} 
\newcommand{\eea}{\end{eqnarray}} 
\newcommand{\mbb}{\mathbb}

\def\IR{\relax{\rm I\kern-.18em R}}

\def\IP{\relax{\rm I\kern-.18em P}} 
\def\inbar{\vrule height1.5ex width.4pt depth0pt} 
\def\IC{\relax\,\hbox{$\inbar\kern-.3em{\rm C}$}}

\def\K3{{\bf K3}}

\makeatletter
\def\Ddots{\mathinner{\mkern1mu\raise\p@
\vbox{\kern7\p@\hbox{.}}\mkern2mu
\raise3\p@\hbox{.}\mkern2mu\raise5\p@\hbox{.}\mkern1mu}}
\makeatother
\begin{document}

%%%%%%%%%%%%%%%%%%%%%%%%%%%%%%%%%%%%%%%%%%%%%%% 
%%%%%%%%%%%%%%%%%%%%%%%%%%%%%%%%%%%%%%%%%%%%%%% 
%%%%%%%%%%%%%%%%%%%%%%%%%%%%%%%%%%%%%%%%%%%%%%% 
%%%%%%%%%%%%%%%%%%%%%%%%%%%%%%%%%%%%%%%%%%%%%%% 
%%%%%%%%%%%%%%%%%%%%%%%%%%%%%%%%%%%%%%%%%%%%%%% 
%%%%%%%%%%%%%%%%%%%%%%%%%%%%%%%%%%%%%%%%%%%%%%% 
%%%%%%%%%%%%%%%%%%%%%%%%%%%%%%%%%%%%%%%%%%%%%%% 
%%%%%%%%%%%%%%%%%%%%%%%%%%%%%%%%%%%%%%%%%%%%%%% 

\vspace*{-1.5cm} 
\begin{flushright} 
  {\small 
  MPP-2008-19 
  } 
\end{flushright}

\vspace{3.2cm} 
\begin{center} 
  {\LARGE 
  Power Towers of String Instantons    
%\\[0.3cm]
 for N=1 Vacua 
  } 
\end{center}

\vspace{1.25cm} 
\begin{center} 
  {\small 
  Ralph~Blumenhagen,  Maximilian Schmidt-Sommerfeld \\ 
  } 
\end{center}

\vspace{0.1cm} 
\begin{center} 
  \emph{
  Max-Planck-Institut f\"ur Physik, F\"ohringer Ring 6, \\ 
  80805 M\"unchen, Germany } 
\end{center}

\vspace{-0.1cm} 
\begin{center} 
  \tt{ 
  blumenha, pumuckl@mppmu.mpg.de \\ 
  } 
\end{center}

\vspace{1.5cm} 
\begin{abstract} 
\noindent  
We provide  arguments for the existence of
novel hereinafter called poly-instanton
corrections to holomorphic couplings in four-dimensional
N=1 supersymmetric string compactifications.
After refining quantitatively the D-brane instanton calculus 
for corrections to the gauge kinetic function,
we explicitly apply it to the Type I toroidal  orbifold
defined in arXiv:0710.3080 and compare the results to the proposed
heterotic S-dual model. 
This leads us to the intriguing conclusion
that N=1 string vacua feature 
a  power tower like  proliferation of instanton corrections.
\end{abstract}

\thispagestyle{empty} 
\clearpage 
\tableofcontents

%%%%%%%%%%%%%%%%%%%%%%%%%%%%%%%%%%%%%%%%%%%%%%% 
%%%%%%%%%%%%%%%%%%%%%%%%%%%%%%%%%%%%%%%%%%%%%%% 
%%%%%%%%%%%%%%%%%%%%%%%%%%%%%%%%%%%%%%%%%%%%%%% 
%%%%%%%%%%%%%%%%%%%%%%%%%%%%%%%%%%%%%%%%%%%%%%% 
%%%%%%%%%%%%%%%%%%%%%%%%%%%%%%%%%%%%%%%%%%%%%%% 
%%%%%%%%%%%%%%%%%%%%%%%%%%%%%%%%%%%%%%%%%%%%%%% 
%%%%%%%%%%%%%%%%%%%%%%%%%%%%%%%%%%%%%%%%%%%%%%% 
%%%%%%%%%%%%%%%%%%%%%%%%%%%%%%%%%%%%%%%%%%%%%%% 

\section{Introduction} 
 
Non-perturbative effects not only play a very important
role in field theories but also in string theory.
For this reason they have been studied
in string theory from the early days on (see for instance \cite{Dine:1986zy,Dine:1987bq,Witten:1996bn,Distler:1987ee,Witten:1999eg,Buchbinder:2002ic,Beasley:2005iu}),
where particular attention  was given to instanton corrections to
the superpotential of the four-dimensional $N=1$ supersymmetric effective action, as
these corrections influence the vacuum structure of
the string compactification. 
For four-dimensional models
very powerful non-renormalisation theorems for
holomorphic couplings have been argued for \cite{Dine:1986zy,Dine:1987bq}. 

Historically, these theorems were first derived
for world-sheet instanton corrections to
the holomorphic couplings in  ${N}=1$ compactifications of the
heterotic string.
In this case, the non-renormalisation theorem states 
that 
the superpotential is only corrected by world-sheets of genus zero, i.e.
single isolated instantons  have the topology of a sphere. 
Similarly, the gauge kinetic functions should
only receive corrections from world-sheets
of genus one, i.e. with the topology of a torus.
These theorems are considered to be confirmed both by explicit computations of
gauge threshold corrections for toroidal orbifolds 
\cite{Dixon:1990pc,Mayr:1993mq}, 
as well as by the implications of
the very powerful method of mirror symmetry \cite{Candelas:1990rm}.

Recently, also space-time instantons have been studied
more concretely \cite{Billo:2002hm,Blumenhagen:2006xt,Ibanez:2006da,Florea:2006si,Akerblom:2006hx,Bianchi:2007fx,Cvetic:2007ku,Argurio:2007vqa,Bianchi:2007wy,Ibanez:2007rs,Akerblom:2007uc,Grimm:2007xm,RoblesLlana:2007ae,Blumenhagen:2007bn,Ibanez:2007tu,Petersson:2007sc,Cvetic:2007qj}. 
In view of the fact that we are only endued with 
 a perturbative approach to string theory, one expects that these are much
harder to describe, as they are  non-perturbative in $g_s$.
However, a subset of these  space-time instantons can be described
as D-branes localised in the four-dimensional space-time and wrapping
a cycle of the internal geometry. Indeed  the microscopic description
of such instantons could be made very explicit by employing
the fact that
their fluctuations are described by an open string theory, just as those of space-time filling D-branes.
This allowed a simple determination of the instanton
zero modes \cite{Ganor:1996pe,Green:1997tv} and, generalizing the 
holomorphy arguments from
the heterotic string, the proposal for an instanton calculus
for holomorphic couplings \cite{Blumenhagen:2006xt}. 
One ingredient of the latter is that the one-loop
determinants describing the fluctuations around the instanton to first order
are captured by open string one-loop diagrams
with one boundary on the instanton. In addition,
following the rules for open strings ending on D-branes,
it was pointed out that instantons with so-called charged zero modes
can generate certain charged matter couplings in the
superpotential which are forbidden
perturbatively \cite{Blumenhagen:2006xt,Ibanez:2006da,Abel:2006yk,Blumenhagen:2007zk}.

To summarise, the advanced techniques of mirror symmetry
allowed to compute whole sums of world-sheet instanton
corrections to holomorphic couplings, while
D-brane and open string technology has its strength
in relating  the microscopic
instanton computation to boundary conformal field
theory. 

In this paper, we follow this second strategy and from this vantage
point revisit
the instanton corrections to holomorphic
couplings in ${N}=1$ four-dimensional orientifold
vacua. 
After quantitatively refining the D-brane instanton
calculus for the gauge kinetic function
proposed in \cite{Akerblom:2007uc}, 
we will argue that, in contrast to field theory, in string
theory there exist instanton corrections to
the instanton action, which leads to an infinite  
power tower-like proliferation
of instanton corrections. As we will see,
these iterated instanton corrections can be understood
as multi-instanton corrections involving different
stringy instantons very much in the spirit of
multi-instanton corrections 
responsible for the correct behaviour of the 
superpotential along lines of marginal stability
\cite{GarciaEtxebarria:2007zv}. 
We would like to emphasise that these effects are not
equivalent to ordinary multi-instantons in field theory, 
which correspond  in string theory to multi D-instantons
wrapping the same internal cycle and placed on top of a stack of space-time
filling D-branes. Since in our case the instantons
wrap different cycles, in order  to distinguish them,
we will call them poly-instantons \footnote{Etymologically,
it would actually be more appropriate  to call them multi-instantons
and the instantons wrapping the same cycle poly-instantons.}.

By S-duality, their existence would imply that in the heterotic
picture there are not only genus zero world-sheet instanton corrections
to the superpotential but also poly-instanton corrections
where precisely one world-sheet has genus zero and all others
genus one. We reckon that these corrections can not be obtained from
Polyakov's path integral for a single string world-sheet, 
as they originate from multiple world-sheets and their interactions.
By interactions we do not mean the usual splitting and joining processes
of strings, but terms in the effective action of two fundamental heterotic strings
that only exist when two world-sheets are present
\footnote{This is analogous to the well known
fact, that the effective $U(N)$ gauge theory on a stack of D-branes
contains new interaction terms compared  to  a single D-brane carrying
only an abelian $U(1)$ gauge symmetry.}.
At least, this seems to be  the picture
imposed upon us by  assuming the validity of both S-duality and our D-brane instanton calculus.

This paper is organised as follows:
In section 2 we present our arguments for the existence of
these novel poly-instanton corrections to the holomorphic gauge kinetic 
function and the superpotential.
In addition, on a more technical level, we  
refine the D-brane instanton calculus
for the computation of holomorphic functions.
In particular, we relate all relevant annulus amplitudes
responsible for the absorption of zero modes to amplitudes known from computations of the
gauge threshold corrections and second derivatives thereof.
To test the proposed calculus, in section 3 we work out
explicitly a recently presented heterotic-Type I
S-dual orbifold model \cite{Camara:2007dy}, where in the heterotic 
description the world-sheet instanton corrections can be computed
explicitly. We compare this result to  the 
expectation from the Type I side. Firstly we find
indirect confirmation of  the Type I
instanton calculus and secondly observe that the poly-instantons
are not  included in the heterotic computation.
We conclude with a number of remarks concerning the
generality of these poly-instanton corrections and their
phenomenological implications.

 %%%%%%%%%%%%%%%%%%%%%%%%%%%%%%%%%%%%%%%%%%%%%%% 
%%%%%%%%%%%%%%%%%%%%%%%%%%%%%%%%%%%%%%%%%%%%%%% 
%%%%%%%%%%%%%%%%%%%%%%%%%%%%%%%%%%%%%%%%%%%%%%% 
%%%%%%%%%%%%%%%%%%%%%%%%%%%%%%%%%%%%%%%%%%%%%%% 

\section{Poly-instanton corrections}
 
In this section, we present  an observation
about string instanton corrections to holomorphic couplings
in ${N}=1$ supersymmetric orientifold compactifications,
which to our knowledge has not yet been spelled out explicitly
and which suggests novel instantons corrections.
By duality, we then expect these corrections to exist in
the heterotic setting, for which  most of the early
string instanton arguments were derived \cite{Dine:1986zy,Dine:1987bq,Witten:1996bn,Distler:1987ee}. 
We guess that these corrections were overlooked mainly for the reason that they
are not so straightforward to see there.

For concreteness we consider the Type IIB orientifold
where the orientifold projection is just the world-sheet
parity transformation $\Omega$. This simple orientifold
is usually called the Type I string. 
To break supersymmetry down to ${N}=1$ in four dimensions,
we compactify the Type I string on a Calabi-Yau manifold
and we introduce $D9$-branes (and
their $\Omega$ images), which can be magnetised,  and $D5$-branes to cancel the tadpoles of the
$O9$ and curvature induced $O5$-planes. 
Note that $D9$-branes invariant under $\Omega$ carry
$SO$ Chan-Paton factors  and $D5$-branes $USp$ ones \cite{Gimon:1996rq}.
Conversely, $\Omega$-invariant euclidean $E5$-branes carry $USp$ Chan-Paton labels
and $\Omega$-invariant euclidean $E1$-branes $SO$ ones.

\subsection{Instanton corrections to the gauge kinetic function}

On such a stack of magnetised $D9$- or $D5$-branes we can compute the Wilsonian 
gauge kinetic function $f_a$, which due to holomorphy
has an expansion 
\bea 
\label{gaugenon} 
    f_a=M_a^0\, \mathcal{S}+ \sum_I M^I_a\, \mathcal{T}_I + f_a^{\text{1-loop}}\left( \mathcal{U}_I\right) 
      + \sum_{E1-{\rm inst.}} g(\mathcal{U}_I)\,\,  e^{2\pi i\, a^I\, \mathcal{T}_I} \; ,  
\eea 
where  the coefficients $M_a^I$ depend on the type of brane and the gauge fluxes
turned on.
Note that in particular  the  one-loop correction and the $E1$-instanton
prefactor do not depend  on the complexified K\"ahler moduli 
$\mathcal{T}_I=c_I+i{\rm Vol}(\Gamma_I)/g_s$, but
only on the complex structure moduli $\mathcal{U}_I$.
Moreover, there are no corrections from $E5$-instantons as they carry
$USp$ Chan-Paton labels and therefore do not have the right zero mode structure.

As shown in the T-dual picture of intersecting
D6-brane orientifolds \cite{Akerblom:2007uc}, the $E1$-instantons must be of type
$O(1)$ and must, in addition to the universal four bosonic and two fermionic zero modes related to
broken translation invariance and supersymmetry, carry two further
fermionic zero modes $\mu^\alpha$, which arise from
the two Wilson lines along the  genus one holomorphic curve wrapped by
the $E1$-instanton.   
 
\begin{figure}[ht]
\centering
\hspace{40pt}
\includegraphics[width=0.6\textwidth]{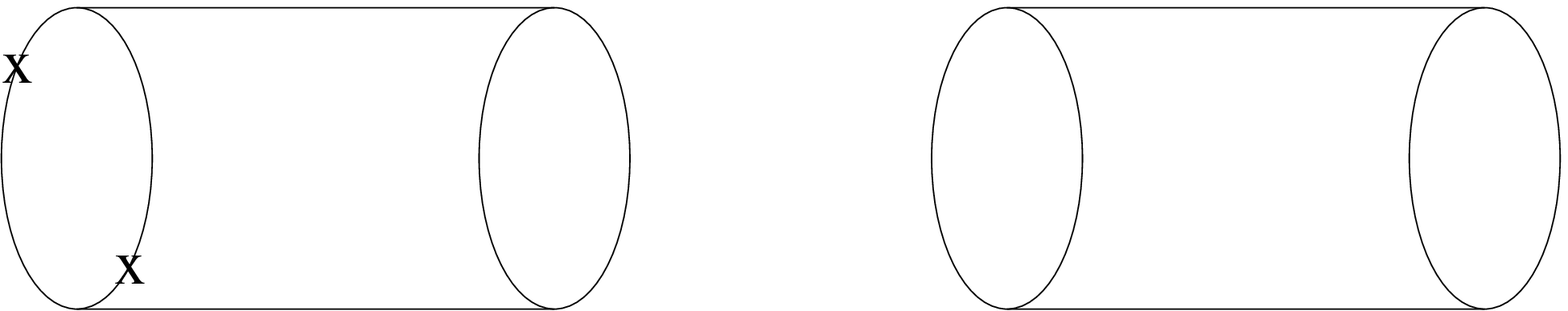}

\begin{picture}(100,1)
\put(-50,38){$D5_r$}
\put(25,38){$D9_b$}
\put(62,38){$=\, {\rm \Re}\Biggl[$}
\put(199,38){$\Biggr]$}
\put(98,38){$E1_r$}
\put(173,38){$D9_b$}
\put(-62,50){$F$}
\put(-26,20){$F$}

\end{picture}

\vspace{-10pt}
\caption{Annulus 1-loop vacuum diagram.\label{figannu}}
\end{figure}

\begin{figure}[ht]
\centering
\hspace{40pt}
\includegraphics[width=0.6\textwidth]{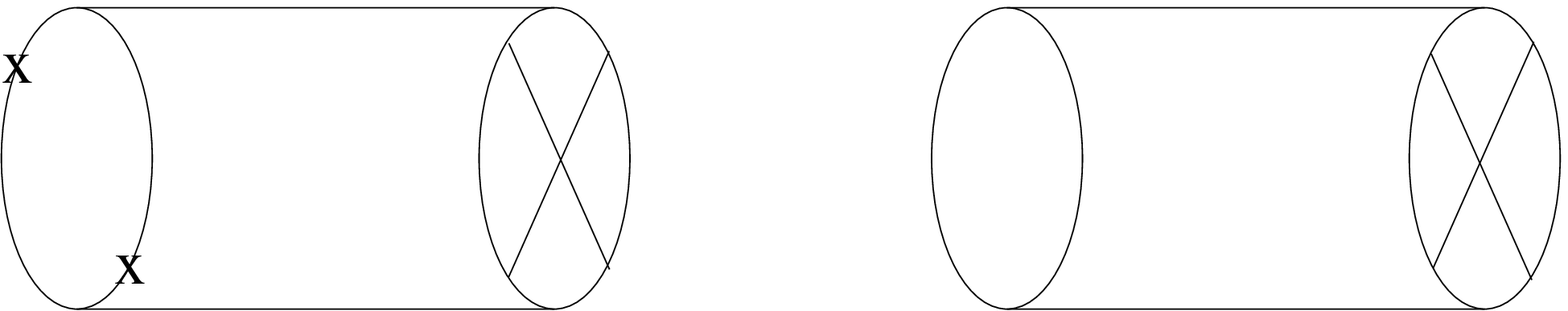}

\begin{picture}(100,1)
\put(-50,38){$D5_r$}
\put(28,38){$O9$}
\put(62,38){$=\, {\rm \Re}\Biggl[$}
\put(199,38){$\Biggr]$}
\put(98,38){$E1_r$}
\put(173,38){$O9$}
\put(-62,50){$F$}
\put(-26,20){$F$}

\end{picture}

\vspace{-10pt}
\caption{M\"obius strip  1-loop vacuum diagram.\label{figmoebi}}
\end{figure}

\begin{figure}[ht]
\centering
\hspace{10pt}
\includegraphics[width=0.4\textwidth]{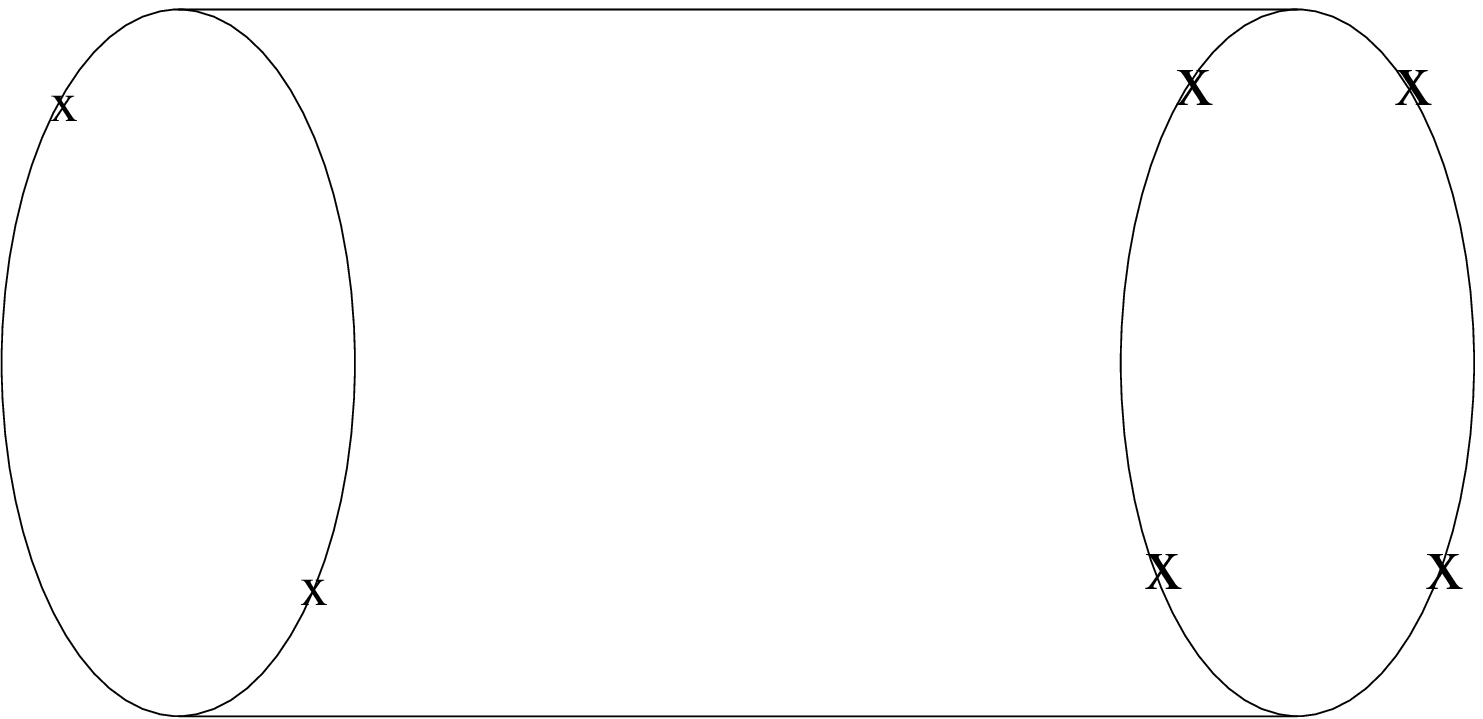}

\begin{picture}(100,1)
\put(-130,50){$A(\underline{D9}_a,E1_r)\ \ =$}
\put(-15,50){$D9_a$}
\put(110,50){$E1_r$}
\put(-36,75){$F_a$}
\put(13,26){$F_a$}
\put(88,75){$\theta^{1\over 2}$}
\put(88,21){$\theta^{1\over 2}$}
\put(140,75){$\mu^{-{1\over 2}}$}
\put(140,23){$\mu^{-{1\over 2}}$}
\end{picture}

\vspace{-10pt}
\caption{Annulus diagram for an $E1$-instanton correction to $f_a$. 
The upper indices give the ghost number of the vertex operators.\label{finst}}
\end{figure}

An instanton with precisely one pair  of  such fermionic zero 
modes  can generate a correction to the $SU(N_a)$ gauge kinetic function
via the instanton correlator
\bea 
\label{gaugekin} 
         \langle F_a(p_1)\, F_a(p_2)  
\rangle_{E1}  
 = \int d^4 x\, d^2\theta \, d^2 \mu\ 
       A(\underline{D9}_a,E1) \, \exp ({-S_{E1}}) \, 
         \, \exp \left(-{Z'_0 (E1)}\right)\,\, , \nonumber 
\eea 
where $S_{E1}$ denotes the tree-level instanton action,
$Z'_0 (E1)$ the one-loop determinants with the zero modes
removed and $A(\underline{D9}_a,E1)$ is the annulus diagram depicted in figure \ref{finst}.
Note  that due to holomorphy, contributions to $f_a$ can reliably 
be computed in the one-loop approximation.
As was first shown in \cite{Abel:2006yk,Akerblom:2006hx}, the one-loop determinants  can be expressed
as the holomorphic part of the M\"obius and the $E1$-$D9$ annulus vacuum amplitudes
\bea
    Z'_0 (E1_r)=\sum_b A({E1}_r,D9_b) + A(E1_r,O9)
\eea
which are related to one-loop  gauge threshold corrections for 
fictitious space-time filling $D5$-branes that are described by the same boundary state in the internal CFT
as the instanton\footnote{We will from now on refer to these D-branes, which
are not there in our model(s), but which are useful to establish
relations and clarify our arguments, as fictitious D-branes.}. This relation is
diagrammatically shown in figures \ref{figannu} and
\ref{figmoebi}.

Note that this relation between the instantonic vacuum
diagrams and the gauge threshold corrections for the fictitious
space-time filling $D5$-brane, identical to the $E1$-instanton in the internal CFT, is expected  from the
observation that for the case the $D5$-brane is really there, i.e. the instanton is lying
on top of a $D5$-brane, the $E1$-instanton describes a gauge instanton
whose instanton action is the gauge coupling \cite{Akerblom:2006hx}.

The four fermionic zero modes of the instanton can be absorbed
by an annulus diagram with appropriate insertions.
Indeed,  $A(\underline{D9}_a,E1)$ is the
annulus diagram shown in figure \ref{finst}.

We will argue later that we can express the one-instanton correction
to the gauge kinetic function on a brane $D9_a$ in terms
of the holomorphic parts of the 
gauge threshold diagrams for a fictitious $D5_r$ brane wrapping
the same internal cycle as the $E1_r$ euclidean brane.
We find it very useful to introduce a diagrammatic notation
for the D-brane instanton amplitudes which illuminates
on first sight which open string annulus and M\"obius diagrams
contribute and how the zero modes
are absorbed. 
\bea 
\label{gaugeka} 
    f^{E1_r}_a=\int d^2\theta_r \, d^2\mu_r\  \,\, \anntf{D9_a}{E1_r}\ \, e^{  - S_{E1_r}}
    \,\, \exp\!\left(   -  \sum_b\: \annbb{E1_r}{D9_b} - \annbc{E1_r}{O9}
    \right) \; .
\eea 

Apparently, what builds up in the exponent is the 
tree level instanton action and its one-loop correction.
The tree level instanton action and the tree level gauge kinetic function
on the fictitious $D5_r$ brane are obtained by dimensionally reducing the
DBI action on the same cycle of the internal manifold and therefore equal:
\bea
  S_{E1_r}^{\rm tree} = 2\pi\, f_{D5_r}^{\rm tree} = 
      \frac{8\pi^2}{ \left(g^{\rm tree}_{D5_r}\right)^2} + i \theta_{D5_r}^{\rm tree}
\eea

Furthermore, as we just argued, the one-loop Pfaffian/determinant of the instanton, which
can be interpreted as the one-loop correction to its action, is equal to
the one-loop gauge threshold corrections on the fictitious
$D5_r$ brane
\bea
 \Re \left[ S_{E1_r}^{\rm 1-loop} \right] = \Re \left[ Z'_0 (E1_r) \right] = 
\frac{8\pi^2}{ \left(g^{\rm 1-loop}_{D5_r}\right)^2}
\eea
and this equality also holds for the holomorphic parts \cite{Akerblom:2007uc}:
\bea
 \mathfrak{Hol} (S_{E1_r}^{\rm 1-loop}) = 2\pi\, f_{D5_r}^{\rm 1-loop}
\eea
In string theory we expect
that the gauge kinetic function on $D5_r$
receives $E1$-instanton corrections, just as that on $D9_a$. Consequently,
if, as one would expect, the aforementioned equality which holds at tree and one-loop level is true exactly,
the $E1_r$ instanton action must receive
the same instanton corrections.
\bea
 S_{E1_r}^{np} \overset{?}{\neq} 0 \qquad \qquad \mathfrak{Hol} (S_{E1_r}^{np}) \overset{?}{=} 2\pi \, f_{D5_r}^{np}
\eea
The latter originate from $E1_s$-branes wrapping
different holomorphic curves of genus one. 
By including these corrections, we obtain an expression
like
\bea 
\label{gaugekb} 
    f_a=\int d^2\theta_r \, d^2\mu_r\   \anntf{D9_a}{E1_r}\
    \,\,  e^{- S_{E1_r} -  Z_0'(E1_r) - \sum_s \int d^4 x_{rs}\, d^2\theta_s\, d^2\mu_s\  
   \annsbf{E1_r}{E1_s}\,\,\,
       e^{  - S_{E1_s} -  Z_0'(E1_s)\ldots  }},\hspace{0.3cm} 
\eea 
where, by iteration of our argument, the dots mean
instanton corrections to the instanton action of $E1_s$.
Here we have already performed a change of integration variables from the
bosonic zero modes $x^\mu_r$ and $x^\mu_s$ to their difference and sum.
This sum, the center of mass position, appears in  the $\int d^4 x\, f_a$ measure factor
and the relative position $x_{rs}=(x_r-x_s)$ is to be integrated over in
(\ref{gaugekb}). 

\noindent
By restricting to a single $E1_s$ instanton (i.e. no summation over $s$ in
\eqref{gaugekb})
and expanding the exponential we can write
\bea 
\label{multiinst} 
 f_a\!\!&=&\!\!\int d^2\theta_r \,d^2\mu_r\  \anntf{D9_a}{E1_r}\,e^{- S_{E1_r}}\, e^{-Z_0'(E1_r)}\,
       \times \\
   &&\phantom{aaaaa} \left[ \sum_{n=0}^\infty 
     \int d^{4n} x_{rs}\, \, d^{2n}\theta_s\, d^{2n}\mu_s\
    {(-1)^n\over n!}\,
       \left( \annbf{E1_r}{E1_s} \right)^n
       e^{- n\, S_{E1_s}}\, e^{- n\, Z_0'(E1_s)}\, \right] \nonumber\\
&=&\int d^2\theta_r \,d^2\mu_r\  \anntf{D9_a}{E1_r}\,\,
       e^{- S_{E1_r}} \, e^{-Z_0'(E1_r)} - \nonumber \\
  && \hspace{-0.5cm} \int d^4 x_{rs}\, d^2\theta_r \, d^2\theta_s\, d^2\mu_r\, d^2\mu_s\
      \anntf{D9_a}{E1_r}\,  \annbf{E1_r}{E1_s}\,\,
         e^{-Z_0'(E1_r)- Z_0'(E1_s)}\, e^{- S_{E1_r}- S_{E1_s}}+\ldots \nonumber 
\eea 
which reveals that these instanton corrections
to the instanton action show up as multiple instanton corrections
to more physical quantities, such as
the gauge kinetic function on the $D9_a$ branes, or, as another example that
we will encounter later on, the superpotential. 
However, these corrections are of a different nature
than multi-instanton contributions in field theory or
standard heterotic multi world-sheet instantons, as
they involve more than one type of $E1$ instantons. We will therefore call them poly-instanton corrections.
Indeed, as we will see later, one gets $\annsbf{E1_r\,}{\, E1_r}=0$ if both
boundaries are the same $E1$
so that an annulus with boundaries on two $E1$-instantons can only absorb the fermionic zero modes
if the boundaries are on different instantons.

Now that we have related instanton corrections to the instanton action to
poly-instanton amplitudes, we would like to compute these
poly-instanton amplitudes to see whether our expectations are fulfilled.
Clearly, in a poly-instanton sector we get many more zero modes
all of which have to be soaked up to yield a non-zero result.
The equation (\ref{multiinst}) already tells us how this 
should happen. For concreteness, let us discuss the
two instanton sector. From the expansion of (\ref{gaugekb})
it is obvious how the zero mode absorption should work
for the higher order terms.

The instanton $E1_r$ corrects the gauge kinetic function on
$D9_a$, so it must be of type $O(1)$ and carry two Goldstino zero modes 
$\theta^\alpha_r$ and two Wilson line modulini zero modes $\mu^\alpha_r$.
The instanton $E1_s$ corrects the instanton action $E1_r$ or,
equivalently (at least we expect so), the gauge
kinetic function on the fictitious $D5_r$ brane, so it must also
be of type $O(1)$ and be endowed with the same zero modes, i.e. two 
$\theta^\alpha_s$'s and two $\mu^\alpha_s$'s.
We require that there are no further zero modes from open strings 
stretched between $E1_r$ and $E1_s$. If such modes were present,
there would be charged zero modes in the $E1_s$-$D5_r$ sector
and the instanton $E1_s$ would not correct the gauge kinetic function
on the fictitious $D5_r$, so we would not expect it to correct
the instanton action of $E1_r$. Consequently,
there are eight fermionic zero modes that need to be saturated in
this two-instanton amplitude.
Four of them, the $\theta^\alpha_r$'s and the $\mu^\alpha_r$'s, can be soaked
up by the amplitude $\annstf{D9_a\,}{\,E1_r}$ (see figure \ref{finst}) and the remaining
ones by the pure instanton diagram $\annsbf{E1_r\,}{\,E1_s}$.
Of course, the role of the two instantons can  be exchanged and
eventually one has to sum over all possibilities of distributing the
fermionic zero modes on different annuli. One has to make sure though that
the whole instanton amplitude is connected from the space-time point of view, i.e.
that the instanton amplitude cannot be factorised into a product
of lower order amplitudes.
In the next section we clarify what happens to the additional
bosonic zero modes $x^\mu_{r,s}$. 

Note that at third order (i.e. for $n=2$ in \eqref{multiinst}) the zero mode absorption requires
the product of three diagrams
\bea
       \anntf{D9_a}{E1_r}\  \annbf{E1_r}{E1_{s}}\   \annbf{E1_r}{E1_{s}}\,\,
\eea
i.e. all additional $\theta$ and $\mu$ zero modes are absorbed
by annulus diagrams with the $E1_r$ instanton on the
empty boundary. In this sector the instanton $E1_s$ and
the zero mode absorption amplitude $\annsbf{E1_r\,}{\,E1_{s}}$ appears twice,
so that one has to insert the usual combinatorial factor $1/2!$.
Extra zero modes appear when the positions of
the two instantons $E1_{s1}$ and $E1_{s2}$ are identical, which
however does not influence the zero mode absorption
amplitudes. Since the sector with these additional zero modes is bose-fermi
degenerate, also the one-loop determinants 
$\exp{\bigl(\ \annsbb{E1_{s1}\ }{\ E1_{s2}} \ \bigr)}$ are not divergent,
whether one includes the zero modes in them or not.
Therefore, it seems to be a fair procedure to evaluate first 
the three and also higher order
amplitudes in the region of instanton moduli space
where the instantons are separated by a finite distance in the four-dimensional spacetime.
Then, if on  the subspace, where the instantons are coincident
in the four dimensional spacetime,
no sources of singularities
appear in the integrand,
one can reliably  take that result. It would be interesting
to honestly perform
the instanton computation with the zero modes for coinciding
instantons included, but this is beyond the scope of this paper.

Let us summarise: Due to the existence of $O(1)$ $E1$-brane instanton corrections
to the gauge kinetic function on both the $D9$ and $D5$ branes
and the fact that the $E1$ instanton action is related to the gauge kinetic
function on fictitious $D5$-branes being identical to the $E1$
instanton in the internal space, one gets instanton corrections to instanton actions
for ${N}=1$ orientifold vacua.
As an immediate consequence this leads to a
proliferation of possible instanton corrections to the
holomorphic gauge kinetic function. These additional corrections
can be understood  as novel stringy poly-instanton corrections
for which part of the zero modes are soaked up
among the different instantons themselves.

Note that if the action of a genus one instanton $A$ receives
corrections from a genus one instanton $B$, then also the opposite is true.
As a consequence one gets, already when there are only two separate instantons with
the appropriate zero mode structure present, an iterative structure
of mutual instanton corrections leading schematically to an infinite power tower 
like
\bea
 \label{pentenried}
    \delta f_a=  e^{-S_A-e^{-S_B-e^{-S_A-{e^{-S_B}}^{\Ddots}}}} \, .
\eea
The generalisation of this expression to the case that more than two instantons
contribute is obvious \footnote{In this paper we are not concerned
with the convergences of such infinite power towers, a question
which deserves investigation. Note that sometimes such  self-similar,
iterated series show  a fractal structure.}.

\subsection{Instanton zero mode absorption}

Except for the instantonic one-loop vacuum diagrams the main
building blocks in the instanton amplitudes are the zero mode
absorption diagrams $\annstf{D9}{E1}$ and $\annsbf{E1}{E1}$. 
These are annulus diagrams with different boundaries
and four fermions inserted on the $E1$ boundary and
are therefore not so straightforward to compute using conformal field
theory methods. We will now argue that by N=1
space-time supersymmetry these diagrams are related
to diagrams with boson vertex operators inserted, which are comparably easy
to compute.

\begin{figure}[ht]
\centering
\hspace{10pt}
\includegraphics[width=0.4\textwidth]{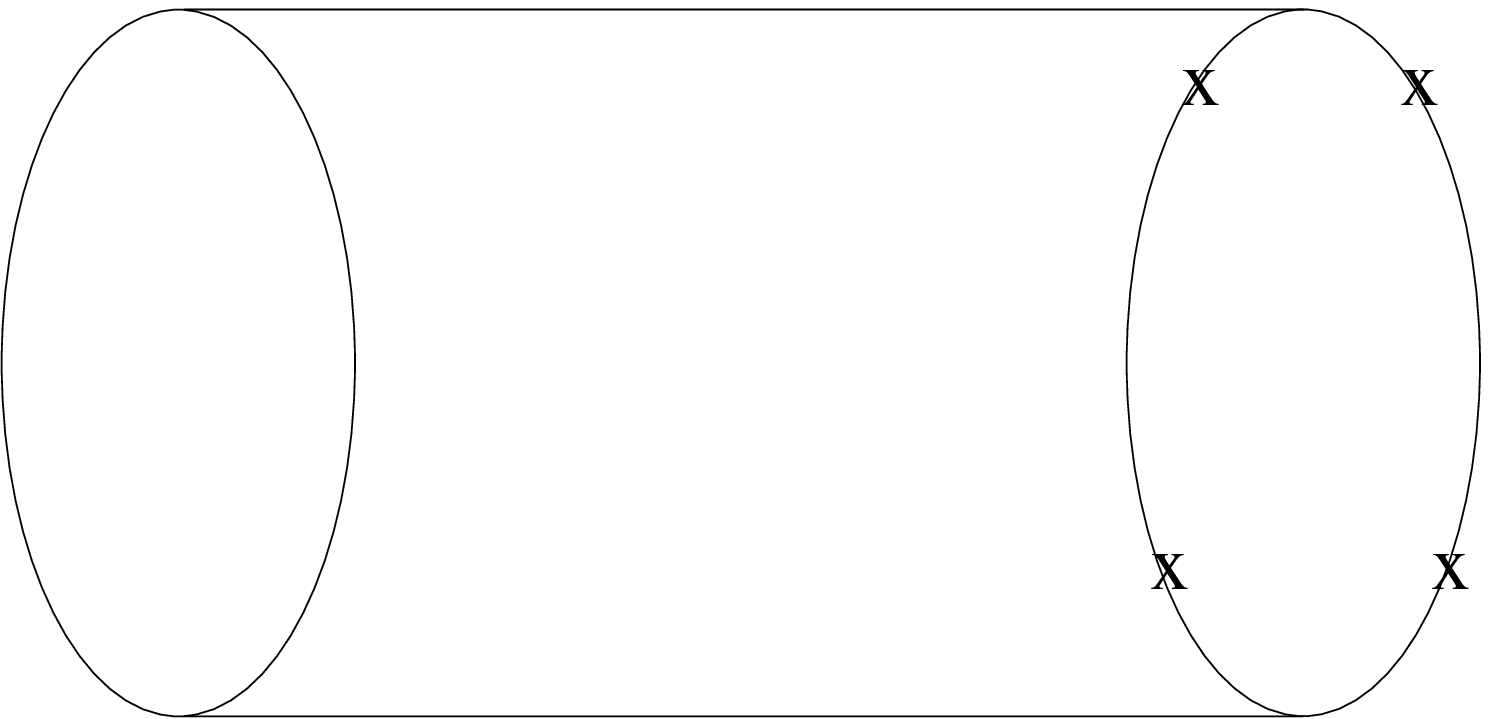}

\begin{picture}(100,1)
\put(-15,52){$D5_r$}
\put(110,52){$D5_s$}
\put(88,76){$\lambda$}
\put(88,26){$\lambda$}
\put(140,78){$\mu$}
\put(140,28){$\mu$}
\end{picture}

\vspace{-10pt}
\caption{Annulus $D5$-brane diagram with four fermion insertions.\label{fiinst2}}
\end{figure}

By ``T-duality'' in the four non-compact directions we expect
that $\annsbf{E1}{E1}$  is  related to the diagram shown in figure \ref{fiinst2}.

\noindent
For $D5$ branes the $\theta$'s become gauginos inside a
vector supermultiplet, usually denoted as
 $\lambda$, and the $\mu$'s are the
Wilson-line modulini inside a chiral multiplet $M=m+\theta \mu$.
Then this coupling  arises from the ${N}=1$
gauge kinetic term 
\bea
      \int d^4x\, d^2\theta\,\, f_s(m+\theta\mu) \, \, {\cal W}^2
\eea
and is the second derivative of the one-loop
correction to the gauge kinetic function with respect
to the Wilson line chiral supermultiplet, evaluated at $m=m_0$, where $m_0$ is
the Wilson line carried by the $O(1)$ instanton.
Therefore, from this supersymmetric Ward identity 
we expect a relation like
\bea
   \annbf{E1_r}{E1_s} \sim {\partial^2\over \partial {m}^2}
\annbt{D5_r}{D5_s}\biggr\vert_{m=m_0}\, ,
\label{tittmoning}
\eea
where the two vertex operators inserted on the boundary of $D5_s$ are
gauge boson vertex operators. The amplitude in \eqref{tittmoning} is
thus the second derivative of a gauge threshold correction amplitude.

However, there is a subtlety concerning
the bosonic zero modes $x^\mu_r$ and $x^\mu_s$.
The annulus diagram with boundaries on two $D5$-branes that is relevant for
the one-loop threshold corrections
has an extra $1/t^2$ factor in the annulus measure compared
to the case of two $E1$-branes. This factor stems from the integration over the
four-dimensional  momenta of the $D5$-$D5$ strings.
On the other hand the open string spectrum between two euclidean
branes contains an extra $\exp({-\pi t (x_r-x_s)^2})$  factor, if
the branes are localised at different positions in the four-dimensional
non-compact space. However, in the two instanton sector we
have to integrate over the relative distance $x_{rs}=x_r-x_s$
which yields precisely
\bea
             \int_{R^4} d^4 x_{rs}\   \ e^{-\pi t x_{rs}^2}= {1\over t^2}\; ,
\eea
such that after integrating over $x_{rs}$ we find that the amplitudes
in \eqref{tittmoning} are identical (up to possible normalisation factors).

Note that the tadpole divergence for $t\to 0$ that one encounters in the 
threshold computation
is, in the case of the $E1$-instantons, not due to massless tadpoles
but comes from the integration over the non-compact
relative distance between the $E1$-branes. 
From all this we conclude that
the correct identification between the holomorphic piece in 
the four-zero mode absorption
$E1$ amplitude and the second derivative of a corresponding gauge
threshold correction is
\bea
\label{strangef}
   \Re \left[ \int d^4 x_{rs} d^2\theta_s d^2\mu_s \  \annbf{E1_r}{E1_s} \right]
      ={\partial^2\over \partial {m}^2}
 \ \annbt{D5_r}{D5_s}\biggr\vert_{m=m_0} \; .
\eea
Since the tadpole divergence in the threshold correction
is not Wilson line dependent, this gives a finite
result. Moreover, for identical $E1$ branes, the
one-loop thresholds are not Wilson line dependent and
therefore, as claimed earlier, $\annsbf{E1_r\,}{\,E1_r}=0$.

Finally we need the zero mode
absorption amplitude between the $D9$-branes and
an $E1$-instanton. In order to determine it, we recall
that, in an annulus diagram, a boundary on a $D9$-brane with
two gauge boson vertex operators inserted can be replaced
with a boundary on an $E5$-instanton, which is localised in
the four-dimensional space-time and is described by the
same boundary state in the internal CFT as the $D9$-brane.
This relation was proven for the case that no vertex operators
are inserted on the other boundary and argued to be true generally
\cite{Billo:2007sw}. The $E5$-$E1$ diagram with four vertex operators
inserted can be related to gauge threshold diagrams
just as the $E1$-$E1$ diagrams and we finally find

\bea
\label{strangeg}
   \Re \left[ \int d^2\theta_r d^2\mu_r \  \anntf{D9}{E1_r} \right]
      ={\partial^2\over \partial {m}^2}
 \ \annbt{D9}{D5_r}\biggr\vert_{m=m_0} \; 
\eea
so that by this line of arguments we have arrived at the conclusion
that for the computation of poly-instanton corrections
to the gauge kinetic function on some $D9$-branes, all
building blocks are related directly or via second derivatives
with respect to the Wilson-line moduli to the holomorphic pieces
of one-loop gauge 
threshold corrections among pairs of (partially fictitious)
space-time filling $D9$ and $D5$-branes.
As required by holomorphy, all building blocks appear
already in the one-loop approximation.
Once we know all these building blocks, the computation
of poly-instanton corrections becomes a combinatorial exercise.

\subsection{Poly-instanton corrections to the superpotential}
\label{secsuperpot}

The arguments we gave for poly-instanton corrections
to the holomorphic gauge coupling directly carry over
to instanton corrections to the superpotential.
Here standard non-renor\-mal\-isa\-tion theorems state that
the superpotential (possibly depending on charged matter fields $\Phi_i$) has the following form
\bea 
\label{supernon} 
       W=W_{\rm tree} + \sum_{E1-{\rm inst.}} \prod_i \Phi_i \ g(\mathcal{U}_I)\,\,  e^{2\pi i\, a^I\, \mathcal{T}_I}   
\eea 
i.e. beyond tree-level there can only be non-perturbative  
contributions from  $E1$-brane instantons of genus zero (for gauge instantons
also $E5$ instantons are possible). 
This is just  S-dual to   world-sheet instantons
for the heterotic string, where only world-sheets with the
topology of the sphere contribute.

For a simple situation, let us argue that we expect also poly-instanton corrections
to the superpotential. Suppose the gauge coupling on some
D-brane receives beyond one-loop also D-instanton corrections:
\bea
 \frac{1}{g^2_{\rm full}} = \frac{1}{g^2_{\rm tree}} + \frac{1}{g^2_{\rm 1-loop}} + \frac{1}{g^2_{np}}
\eea
Now we consider the four-dimensional low-energy effective field theory on
this D-brane. Let us assume that it is such that an ADS-type superpotential
is dynamically generated by a gauge instanton. The superpotential
will then look like
\bea
 W_{ADS} =  \frac{1}{\det \Phi \bar{\Phi}} \,
             \exp\left({\textstyle -\frac{8\pi^2}{g^2_{\rm full}}}\right)=
            \frac{1}{\det \Phi \bar{\Phi}} \,
      \exp \left({\textstyle -\frac{8\pi^2}{g^2_{\rm tree}} - \frac{8\pi^2}{g^2_{\rm 1-loop}}- \frac{8\pi^2}{g^2_{np}}}\right) .\quad
\eea
It must be possible to derive the superpotential in the full string theory. 
There, it is generated by a D-instanton wrapping the same cycle in the internal space as the D-brane.
Clearly, if the same superpotential is to be generated, the instanton action must
receive the same instanton corrections as the gauge coupling on the D-brane. Analogously to
what we described for the gauge kinetic function, these corrections should arise as
poly-instanton corrections to the superpotential.

Coming back to eq. \eqref{supernon}, 
we therefore expect generally that the instanton action itself can receive
corrections from $E1$-instantons wrapping holomorphic curves of genus one, which,
in terms of corrections to the superpotential, means that the latter receives
poly-instanton contributions of the form
\bea 
\label{multiinstw} 
 W_{\rm inst}&\sim& e^{-Z_0'(E1_r)}\,  e^{- S_{E1_r}}
       +\nonumber \\
       && \annbf{E1_r}{E1_s}\,
         e^{-Z_0'(E1_r)- Z_0'(E1_s)}\, e^{- S_{E1_r}- S_{E1_s}} +\ldots  ,
\eea 
where $E1_r$ wraps a curve of genus zero and $E1_s$ one of genus one.
One might be worried that these contributions of genus one
spoil the celebrated non-renor\-mal\-isa\-tion theorem for
the Type I-heterotic superpotential. 
But since these genus one instantons always appear in poly-instanton
sectors, where precisely one instanton is of genus zero and all others of genus one, the 
dependence on the dilaton superfield $\mathcal{S}$ of these poly-instanton contributions is equal
to that of a one $E1$-instanton contribution, where the instanton wraps a curve of genus zero.
More precisely, the dilaton dependence is characterised by
the sum of the  Euler characteristics of the holomorphic curves
(up to an Einstein-frame induced factor of  $-2$ )
\bea
        \mathcal{S}^{\chi(E1_r)+\sum_s \chi(E1_s)-2}=\mathcal{S}^{\chi(E1_r)-2}=1\, .
\eea
Therefore these poly-instanton corrections are not forbidden
by holomorphy of the superpotential. \footnote{One could even speculate
that there might be contributions from curves of even lower Euler characteristics,
when several spheres are involved, as long as the sum of the Euler characteristics is two.}
Their presence depends on the value of the
coupling $\annsbf{E1_r\,}{\,E1_s}$ and we
do not see any reason why this   should generically
vanish.

It would be interesting to find concrete examples where
these poly-instanton corrections are definitely present.
For the remainder of this paper, we discuss an example where
poly-instanton corrections to the gauge kinetic function
can be computed concretely.

\section{A Heterotic-Type I dual pair}

So far our arguments and equations have been very general
and it could well be that for some unobvious  reason
some of the zero mode absorption annulus diagrams
do vanish. 
Since the appearing gauge threshold corrections can be
explicitly computed for D-branes on  toroidal
orbifolds \cite{Antoniadis:1999ge,Lust:2003ky,Berg:2004ek,Akerblom:2007np,Blumenhagen:2007ip}, in the remainder 
of this paper we will discuss a recently proposed 
heterotic-Type I
S-dual pair of shift orbifolds \cite{Camara:2007dy,dudas} in some detail.
It will turn out to be very  illuminating 
to see how   heterotic-Type I
S-duality works for the instanton corrections of the
gauge kinetic function in this case.

\subsection{The heterotic orbifold}

In \cite{Camara:2007dy} a dual pair of heterotic and Type I models in four dimensions was proposed
based on a freely acting  $\mbb Z_2 \times \mbb Z_2$ orbifold
in four dimensions.
The orbifold is defined on a factorisable torus $T^6=T^2\times T^2\times T^2$ with
the following action of the two $\mbb Z_2$s
\bea
 \label{karajan}
          \Theta:\begin{cases}
                     z_1\to -z_1 \\
                 z_2\to -z_2+{1\over 2} \\
                  z_3\to z_3+{1\over 2} 
                  \end{cases}\quad\quad\
          \Theta':\begin{cases}
                   z_1\to z_1+{1\over 2} \\
                 z_2\to -z_2 \\
                  z_3\to -z_3+{1\over 2} 
                \end{cases}\quad\quad\
                 \Theta'':\begin{cases}
                   z_1\to -z_1+{1\over 2} \\
                 z_2\to z_2+{1\over 2} \\
                  z_3\to -z_3 \; .
                \end{cases}.
\eea
The $\Omega$ orientifold of this model has only one $O9$ plane,
whose tadpole can be cancelled by 32 $D9$-branes yielding the
gauge group $SO(32)$. Moreover, the  $\mbb Z_2 \times \mbb Z_2$ orbifold
action projects out all three complex Wilson lines
of the $D9$-branes, so that the massless spectrum
is that of a pure $SO(32)$ gauge theory coupled to supergravity.

We refer the reader to \cite{Camara:2007dy} for the details of the dual heterotic model,
let us only mention that there
the shift symmetry acts in an asymmetric way  
\bea
    X_L\to X_L+{\pi R\over 2} +{\pi\alpha'\over 2R},\quad
    X_R\to X_R+{\pi R\over 2} -{\pi\alpha'\over 2R},
\eea
i.e. it is a combination of a Kaluza-Klein and a winding shift.
The  orbifold action reduces to the purely geometric one of \eqref{karajan} in
the large volume limit. 

What is important for us,
is that the authors computed the perturbative (in $g_s$) one-loop
gauge threshold corrections for the gauge group $SO(32)$
\bea
      \Re (f_a)=\Im (\mathcal{S}) + \Lambda_a\; ,
\eea
where the running due to massless modes is contained in $\Lambda_a$.
In order to do so, one starts from the following formula
for the one-loop gauge threshold corrections in heterotic compactifications
\cite{Kaplunovsky:1987rp,Kaplunovsky:1992vs}.
\bea
 \Lambda_a = \int_\mathcal{F} \frac{d^2\tau}{\tau_2} \frac{i}{4\pi} \frac{1}{|\eta|^2} \sum_{even (\alpha,\beta)} \partial_{\bar{\tau}}
 \left( \frac{\bar{\vartheta}\genfrac[]{0pt}{}{\alpha}{\beta}}{\bar{\eta}} \right)
{\rm  Tr}_\alpha \left( \left( Q_a^2 - \frac{1}{4\pi\tau_2}\right) (-1)^{\beta \bar{F}} q^H \bar{q}^{\bar{H}} \right)_{int}
 \label{pelle}
\eea
This formula amounts to computing a trace in the Hilbert space of the internal CFT. $Q_a$ is the charge of
a string state under the gauge group $G_a$ under consideration, $\bar{F}$ is the right-moving world-sheet fermion
number and $H$ and $\bar{H}$ are the left- and right-moving world-sheet Hamiltonians. The sum in \eqref{pelle}
runs over the even spin structures of the fermions of the right-moving superstring.

After slightly rearranging the result of \cite{Camara:2007dy}, for later matching with
the Type I side,  the thresholds can be written
in the form\footnote{Note that we have included an extra factor $i$ in the
third term in the first line of \eqref{hetthres} as compared to \cite{Camara:2007dy}
to make \eqref{hetthres} modular invariant and to ensure in particular that
the second and third term in the first line of \eqref{hetthres} transform into each other
through a modular T transformation.}
\bea
\label{hetthres}
    \Lambda&=&\int_{\cal F} {d^2\tau\over \tau_2}\,
     \sum_{i=1}^3 \left(
            {1\over  \eta^2\, \vartheta^2_2 }\, \hat Z_i 
         \left[\begin{matrix} 1\\ 0 \end{matrix} \right]-
      {1\over \eta^2\, \vartheta^2_4}\,  \hat Z_i 
         \left[\begin{matrix} 0\\ 1 \end{matrix} \right]-
         {i\over  \eta^2\, \vartheta^2_3}\,  \hat Z_i 
         \left[\begin{matrix} 1\\ 1 \end{matrix} \right]
 \right)\times\\
   && \quad\quad \sum_{a,b} \Biggl( {\thbb{a}{b}\over \eta}\Biggr)^{16}
         \,\,\,   \Biggl( {\thbb{a}{b}''\over \thbb{a}{b}} + {\pi\over \tau_2} 
         \Biggr) \nonumber 
\eea         
where the last sum runs over the three sectors $(a,b)\in\{ (0,0),(0,1/2),(1/2,0)\}$ and originates
from the $\widehat{SO}(32)_1$ left-moving current algebra. 
Here we have set the normalisation to one, as
throughout our computation we will ignore the overall
moduli-independent normalisation factor. 
In (\ref{hetthres}) one defines
\bea
    \hat Z_i \left[\begin{matrix} h\\ g \end{matrix} \right]
 ={\mathcal{T}^{(i)}_2\over \tau_2} \sum_{n_1,l_1,n_2,l_2}   (-1)^{h\, n_1 + g\, l_1}
\, \exp\left[ 2\pi i\, \det(A)\, \mathcal{T}^{(i)} -{\pi \mathcal{T}^{(i)}_2\over \tau_2 \mathcal{U}^{(i)}_2}\,
    \left\vert (1,\mathcal{U}^{(i)})\, A\, \left(\begin{matrix} \tau\\ -1 \end{matrix}
      \right) \right\vert^2 \right]\nonumber
 \\ 
\label{heteroricKKWsum}
\eea
and the matrix of the Kaluza-Klein and winding modes as
\bea
          A=\left(\begin{matrix} n_1+{g\over 2} & l_1+{h\over 2} \\ n_2 & l_2
            \end{matrix} \right)\; ,
\eea
where for the two Kaluza-Klein sums the 
Poisson resummation formula has already been applied.
Note, that the terms in the first line of (\ref{hetthres})
arise from the right-moving supersymmetric 
sector of the heterotic string and the
terms in the second line from the left-moving bosonic sector.

In \cite{Camara:2007dy} the authors nicely matched the
perturbative (in $\alpha'$) purely complex structure moduli dependent
contributions 
\bea
      \Lambda_{\rm 1-loop}\simeq\log\left({\vartheta_4\over \eta^3}(2\mathcal{U})\right)
\eea
to the gauge kinetic function for this
heterotic - Type I dual pair. They arise in \eqref{heteroricKKWsum} from the
sum over the degenerate orbits with $\det(A)=0$. 
Here we are interested in corrections arising from world-sheet instantons, 
i.e. in terms with $\det(A)\ne 0$.
The questions we would like to answer are: 
\begin{itemize}
\item Can we  quantitatively reproduce
the holomorphic part of the heterotic result on the Type I side in terms of $E1$-brane instanton corrections?

\item Are there poly-instanton contributions on the Type I dual side
     and, if so,  are they also included in the heterotic dual?
\end{itemize}

The first step is to evaluate the integrals in (\ref{hetthres}),
which can be done using the methods introduced
in \cite{Dixon:1990pc,Bachas:1997mc}. 
One starts by unfolding the integral over
the fundamental domain ${\cal F}$ of $SL(2,\mbb Z)$ by 
taking orbits of matrices $A$ under the modular
group. Taking into account that by a modular transformation
the three summands in (\ref{hetthres}) get mutually interchanged, 
it is easy to see that
in each non-degenerate orbit there is precisely one matrix of the form
\bea
\label{matriceskjp}
         A=\left(\begin{matrix} k & j \\ 0 & p
            \end{matrix} \right)\; .
\eea
with $2j,2k,p\in \mbb Z$, $0\le j< k$, but not both $j$ and $k$ integer. Such
an orbit is characterised by $\det(A)=k\cdot p$.
In the following, we will only be concerned with instantons, i.e. $p>0$,
as opposed to anti-instantons, for which $p<0$.
Unfolding the integral then allows one to carry
out the integral, as it becomes an integral over the full upper half $\tau$-plane,
which, following precisely the steps documented
in \cite{Bachas:1997mc}, leads, for fixed $\det(A)$, to the general form
for the holomorphic part
\bea
\label{complhet}
       \Lambda(\vec{\mathcal{U}},\vec{\mathcal{T}})=\sum_{i=1}^3 
        \Lambda(\mathcal{U}_i)\,{1\over \det( A)}\, e^{2\pi i\, \det(A)\, \mathcal{T}^{(i)}}\; .
\eea
For the holomorphic prefactor $\Lambda(\mathcal{U})$ we obtain
\bea
\label{hetoone}
    \Lambda(\mathcal{U})={\cal A}\left[\begin{matrix} 1\\ 0 \end{matrix} \right]
\left( {j+p\mathcal{U}\over k} \right)=
    {(-1)^k\over  \eta^2\, \vartheta^2_2 }\, 
   \sum_{a,b} \Biggl( {\thbb{a}{b}\over \eta}\Biggl)^{\!\! 16}
          \,\,  \Biggl( {\thbb{a}{b}''\over \thbb{a}{b}} \Biggr)\left( {j+p\mathcal{U}\over
                k} \right)
\eea
for $k\in \mbb Z$ and $j\in \mbb Z+{1\over 2}$,
\bea
\label{hetotwo}
    \Lambda(\mathcal{U})={\cal A}\left[\begin{matrix} 0\\ 1 \end{matrix} \right]
\left( {j+p\mathcal{U}\over k} \right)=  -{(-1)^j\over  \eta^2\, \vartheta^2_4 }\, 
   \sum_{a,b} \Biggl( {\thbb{a}{b}\over \eta}\Biggl)^{\!\! 16} \,\,
            \Biggl( {\thbb{a}{b}''\over \thbb{a}{b}} \Biggr)\left( {j+p\mathcal{U}\over
                k} \right)
\eea
for $k\in \mbb Z+{1\over 2}$ and $j\in \mbb Z$ and
\bea
\label{hetothree}
    \Lambda(\mathcal{U})= {\cal A}\left[\begin{matrix} 1\\ 1 \end{matrix} \right]
\left( {j+p\mathcal{U}\over k} \right)={i\,(-1)^{k+j}\over \eta^2\, \vartheta^2_3 }\, 
   \sum_{a,b} \Biggl( {\thbb{a}{b}\over \eta}\Biggr)^{\!\! 16}\,\,
            \Biggl( {\thbb{a}{b}''\over \thbb{a}{b}} \Biggr)\left( {j+p\mathcal{U}\over
                k} \right)
\eea
for $k\in \mbb Z+{1\over 2}$ and $j\in \mbb Z+{1\over 2}$.

The above expressions can be rewritten in terms of the Eisenstein series $E_2$
and $\vartheta$/$\eta$-functions.
Note that these are only the holomorphic parts of the amplitude, whereas
the  integral (\ref{hetthres}) contains also non-holomorphic pieces, e.g. one
originating from the second summand ($\pi/\tau_2$) in the last bracket.
Upon including this term, the above expressions become modular forms
involving $\vartheta$/$\eta$-functions and the modified Eisenstein series
$\hat E_2=E_2-{3\over \pi\tau_2}$.
Here we are only interested in the
holomorphic pieces contained in the Wilsonian gauge kinetic
functions and will therefore mostly neglect all the non-holomorphic
pieces. We will however include this ${\pi/ \tau_2}$ contribution
when concerned with the modular
properties of the expressions obtained.

It  was shown in \cite{Bachas:1997mc} that the sum over the orbits
is nothing else than the  Hecke operator acting  on a modular form,
which defines a new modular form in terms of a finite
sum over a known modular form at shifted arguments.
Due to the  $\mbb Z_2 \times \mbb Z_2$ orbifold action
the modular properties of the thresholds  $\Lambda(\mathcal{U})$ do
change, as the relevant modular group is only the
$\Gamma_2\subset SL(2,\mbb Z)$ subgroup generated
by $\{ T^2,U^2=S\, T^{-2}\, S \}$\cite{Bianchi:2007rb}.
This group is the maximal subgroup of $SL(2,\mbb Z)$ which leaves
the momentum/winding sums in \eqref{hetthres} invariant.
The expression \eqref{complhet} can be considered
as a generalised Hecke operator with respect to the
modular subgroup $\Gamma_2$.

\vspace{0.2cm}

Let us now extract the leading and next to leading order instanton
corrections to see whether they can be understood from the Type I 
dual perspective:

\vspace{0.5cm}
\noindent
{\bf First order  instanton sector}

\vspace{0.1cm}
\noindent

The leading order instantons\footnote{We say that an instanton has  {\it order}
$n$, when it appears with  $\exp(n\, \pi i \mathcal{T})$.}
 have  $\det(A)=1/2$, for which there
exists only one orbit, one representative being
\bea
         A=\left(\begin{matrix} {1\over 2} & 0 \\ 0 & 1
            \end{matrix} \right)\; ,
\eea
so that after some little algebra the instanton correction becomes
\bea
    \Lambda_1 (\mathcal{U},\mathcal{T})&=&2\,{\cal A}\left[\begin{matrix} 0\\ 1 \end{matrix} \right]
\left( {2\mathcal{U}} \right)\, \, e^{\pi i\, \mathcal{T}} \nonumber \\
 &=& \frac{2\,\pi^2}{3} \Bigg[ { e^{\pi i\, \mathcal{T}}\over \eta^4(\mathcal{U})} \, \, 
   \left( {\vartheta_3 \over \eta}(2\mathcal{U})\right)^{\!\! 16}
   \left(\hat E_2+\vartheta_2^4-\vartheta_4^4\right)(2\mathcal{U})+ \label{singlea}\\
 && {e^{\pi i\, \mathcal{T}}\over  \eta^4(\mathcal{U})} \, \, 
   \left( {\vartheta_4 \over \eta}(2\mathcal{U})\right)^{\!\! 16}
   \left(\hat E_2-\vartheta_2^4-\vartheta_3^4\right)(2\mathcal{U})+\label{singleb} \\
 && { e^{\pi i\, \mathcal{T}}\over  \eta^4(\mathcal{U})} \, \, 
   \left( {\vartheta_2 \over \eta}(2\mathcal{U})\right)^{\!\! 16}
   \left( \hat E_2+\vartheta_3^4+\vartheta_4^4\right)(2\mathcal{U})  \Bigg] \label{singlec}
\eea
Our task in the next section will be to identify the
holomorphic contributions in this expression on the
Type I side, not only qualitatively but quantitatively.           
Note that to capture the $\Gamma_2$ modular properties
of this expression we have written $\hat E_2$ instead of the holomorphic
piece $E_2$. The invariance
under $T^2$  is obvious, as the only
pieces in (\ref{singlea})-(\ref{singlec}) transforming non-trivially
are
\bea
 T^2:\vartheta_2(2\mathcal{U})\to -\vartheta_2(2\mathcal{U}), \quad\quad
 T^2:\eta(2\mathcal{U})\to e^{\pi i\over 3 }\, \eta(2\mathcal{U})\; .
\eea
Under $U^2$ the various contributions transform as follows
\bea
\label{ufos}
    U^2:\begin{cases}
          \hat E_2\left({2\mathcal{U}\over 2\mathcal{U}+1}\right)=({\scriptstyle{2\mathcal{U}+1}})^2\,\,
          \hat E_2 (2\mathcal{U})\\
          \eta\left({2\mathcal{U}\over 2\mathcal{U}+1}\right)=e^{-{\pi i\over
            12}}\,\sqrt{\scriptstyle{2\mathcal{U}+1}}\,\,\eta(2\mathcal{U})\\
          \vartheta_4\left({2\mathcal{U}\over 2\mathcal{U}+1}\right)=e^{-{\pi i\over 4}}\,\sqrt{\scriptstyle{2\mathcal{U}+1}}\,\,
          \vartheta_4 (2\mathcal{U})\\
          \vartheta_3\left({2\mathcal{U}\over 2\mathcal{U}+1}\right)=\sqrt{\scriptstyle{2\mathcal{U}+1}}\,\,
          \vartheta_2 (2\mathcal{U})\\
          \vartheta_2\left({2\mathcal{U}\over 2\mathcal{U}+1}\right)=\sqrt{\scriptstyle{2\mathcal{U}+1}}\,\,
          \vartheta_3 (2\mathcal{U})\\
         \end{cases}
\eea
so that (\ref{singleb}) is invariant and (\ref{singlea}) and 
(\ref{singlec}) are exchanged.
Therefore,  $\Lambda_1 (\mathcal{U},\mathcal{T})$ is indeed invariant under
the modular group $\Gamma_2$ acting on the complex structure modulus
$\mathcal{U}$.

\vspace{0.3cm}
\noindent
{\bf Second  order  instanton sector}

\vspace{0.1cm}
\noindent
The next to leading order instantons have  $\det(A)=1$, for which there
exist two orbits. We choose the representatives
\bea
\label{Han96}
         A=\left(\begin{matrix} {1\over 2} & 0 \\ 0 & 2
            \end{matrix} \right),\quad\quad
          A=\left(\begin{matrix} 1& {1\over 2}  \\ 0 & 1
            \end{matrix} \right)\;          
\eea
leading to 
\bea
    \Lambda_2 (\mathcal{U},\mathcal{T})&=&{\cal A}\left[\begin{matrix} 0\\ 1 \end{matrix} \right]
\left( {4\mathcal{U}} \right)\, \, e^{2\pi i\, \mathcal{T}} +
   {\cal A}\left[\begin{matrix} 1\\ 0 \end{matrix} \right]
\left( {{1\over 2}+\mathcal{U}} \right)\, \, e^{2\pi i\, \mathcal{T}}
    \nonumber \\
 &=& \frac{\pi^2}{3} \Bigg[ { e^{2\pi i\, \mathcal{T}}\over \eta^4(2\mathcal{U})} \, \, 
   \left( {\vartheta_3 \over \eta}(4\mathcal{U})\right)^{\!\! 16}
   \left(\hat E_2+\vartheta_2^4-\vartheta_4^4\right)(4\mathcal{U})+ \\
 &&\phantom{aaa} {e^{2\pi i\, \mathcal{T}}\over  \eta^4(2\mathcal{U})} \, \, 
   \left( {\vartheta_4 \over \eta}(4\mathcal{U})\right)^{\!\! 16}
   \left(\hat E_2-\vartheta_2^4-\vartheta_3^4\right)(4\mathcal{U})+ \label{wichta}\\
 && \phantom{aaa} { e^{2\pi i\, \mathcal{T}}\over  \eta^4(2\mathcal{U})} \, \, 
   \left( {\vartheta_2 \over \eta}(4\mathcal{U})\right)^{\!\! 16}
   \left(\hat E_2+\vartheta_3^4+\vartheta_4^4\right)(4\mathcal{U}) \Bigg] + \\
 && \hspace{-1.7cm} \frac{\pi^2 e^{-\pi i/3}}{12} \Bigg[ { e^{2\pi i\, \mathcal{T}}\over \eta^4(2\mathcal{U})} \, \, 
   \left( {\vartheta_3 \over \eta}({\textstyle{1\over 2}}+\mathcal{U})\right)^{\!\!16}
   \left(\hat E_2+\vartheta_2^4-\vartheta_4^4\right)({\textstyle{1\over 2}}+\mathcal{U})+ \\
 && {e^{2\pi i\, \mathcal{T}}\over  \eta^4(2\mathcal{U})} \, \, 
   \left( {\vartheta_4 \over \eta}({\textstyle{1\over 2}}+\mathcal{U})\right)^{\!\! 16}
   \left(\hat E_2-\vartheta_2^4-\vartheta_3^4\right)({\textstyle{1\over 2}}+\mathcal{U})+ \\
  && { e^{2\pi i\, \mathcal{T}}\over  \eta^4(2\mathcal{U})} \, \, 
   \left( {\vartheta_2 \over \eta}({\textstyle{1\over 2}}+\mathcal{U})\right)^{\!\! 16}
   \left(\hat E_2+\vartheta_3^4+\vartheta_4^4\right)({\textstyle{1\over 2}}+\mathcal{U}) \Bigg]
   \label{wichb}
\eea
One can show that this expression is invariant under
the modular subgroup $\Gamma_2$, where in particular
under $U^2$ the two orbits in (\ref{Han96}) get exchanged.

Of course, both in the  first and the second order  instanton sector 
(in fact in any), we get all terms for all three two-tori.
\subsection{The Type I dual}

These heterotic world-sheet instanton
corrections are expected to be S-dual to $E1$-instanton corrections in the
Type I model.
Recall that here we have the $O9$-plane and
32 $D9$-branes, which, in order to get the S-dual of the heterotic model,
must have trivial Wilson lines
along the six one-cycles of $T^6$.

Let us start with the single instanton corrections
to the gauge kinetic function. The contributing
instantons must have the right zero mode structure.
First of all they must be $O(1)$ instantons, which
is guaranteed for $E1$-branes wrapping one of the
three $T^2$ factors with or without a Wilson line turned
on, which is invariant under $\Omega$.
Ignoring first the $\mbb Z_2 \times \mbb Z_2$ 
 orbifold action this is satisfied 
for discrete Wilson lines $(\alpha,\beta)\in\{ (0,0),(0,{1\over 2}),({1\over
  2},0),({1\over 2},{1\over 2}) \}$
along the fundamental one-cycles of the $T^2$ the brane wraps.
However, for trivial Wilson lines $(0,0)$ extra charged zero modes
appear from open strings stretched between the $E1$ and the $D9$ branes.
Moreover, one of the three $\mbb Z_2$s acts on the torus $T^2$ only
by a $\mbb Z_2$ shift $x\to x+{1\over 2}$. 
Taking into account that this shift acts like $(-1)^m$ on
the KK modes, we conclude that a Wilson line  $({1\over 2},.)$ is not allowed
whereas  an odd  Wilson line $(1,.)$ is non-trivial. 

\begin{figure}[ht]
\centering
\hspace{10pt}
\includegraphics[width=0.7\textwidth]{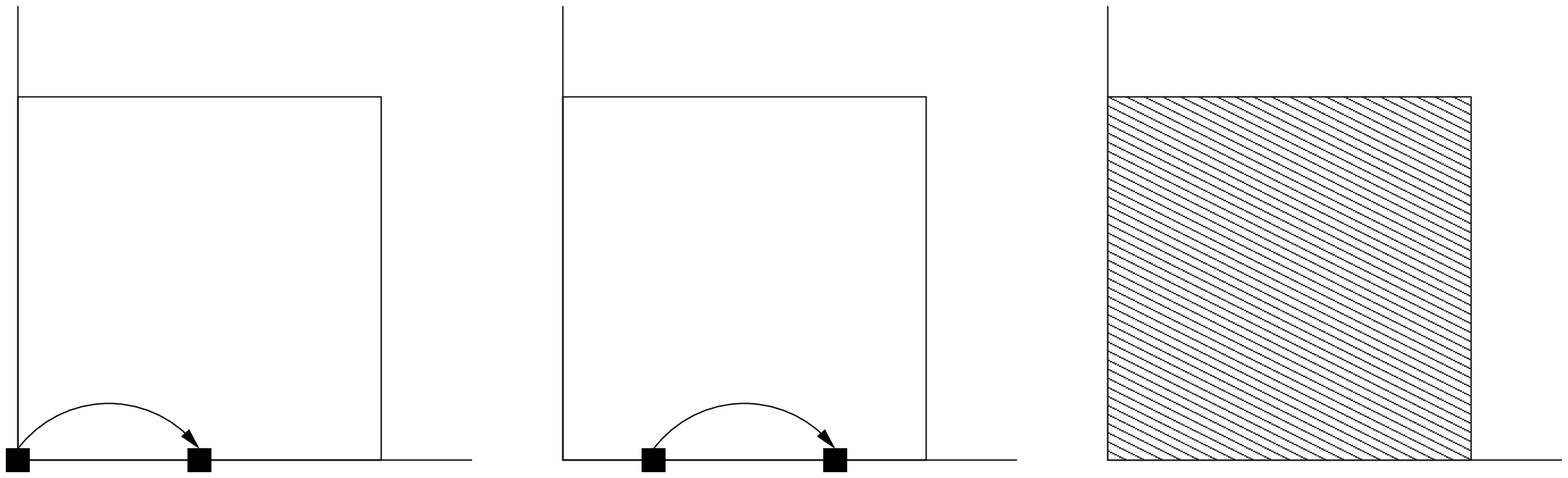}

\begin{picture}(100,1)
\put(-80,32){$\Theta',\Theta''$}
\put(35,32){$\Theta',\Theta''$}
\put(140,92){$T^2_{(3)}$}
\put(39,92){$T^2_{(2)}$}
\put(-62,92){$T^2_{(1)}$}
\put(111,106){$y_{3}$}
\put(10,106){$y_{2}$}
\put(-91,106){$y_{1}$}
\put(202,15){$x_{3}$}
\put(100,15){$x_{2}$}
\put(0,15){$x_{1}$}

\end{picture}

\vspace{-10pt}
\caption{Position of a single $O(1)$ instanton: This $E1$ wraps the third
$T^2$ and is localised on the fixed points $(x_1,y_1)=(0,0)+(1/2,0)$,
$(x_2,y_2)=(1/4,0)+(3/4,0)$ of $\Theta$
 .\label{poseeins}}
\end{figure}

Therefore, for each $T^2$ of the $\mbb Z_2 \times \mbb Z_2$ shift orbifold we have 
three candidate $O(1)$ instantons with discrete Wilson lines
\bea
       E1^{(i)}_{(0,{\textstyle {1\over 2}})}, \quad\quad E1^{(i)}_{(1,0)},
      \quad\quad E1^{(i)}_{(1,{\textstyle {1\over 2}})} \quad\quad i=1,2,3\; .
\eea
In order for them to contribute to the gauge kinetic function,
they must also have precisely the two modulino zero modes
$\mu^\alpha$, i.e. they must be rigid along
the other two $T^2$ factors transverse to the instantons.
But this can be arranged by placing them on the
four possible pairs of fixed points of the $\mbb Z_2$ which acts by a shift
along the $T^2$ wrapped by the instanton like shown in figure \ref{poseeins}.

Therefore, altogether we found $3\cdot 3\cdot 4=36$   $O(1)$ instantons
which have the right zero mode structure to yield
a single instanton correction to the gauge kinetic function
of the pure $SO(32)$ super Yang-Mills theory on the $D9$-branes.
For reasons that will become clear later, let us denote these 
instantons by
\bea
         E1^{i,k}_2, \quad E1^{i,k}_3,\quad E1^{i,k}_4
\eea
with $i=1,2,3$ denoting the $T^2$ factor wrapped by the instanton,
$k=1,2,3,4$ denoting the pair of fixed points where the
instanton is located on the two remaining $T^2$ factors and
finally the lower index denoting the discrete Wilson line
turned on along $T^2_i$ with 
$E1_2=E1_{(0,{\textstyle {1\over 2}})}, E1_4=E1_{(1,0)},E1_3=E1_{(1,{\textstyle {1\over 2}})}$.

\subsection{The one-instanton sector}

According to the formula (\ref{gaugeka}) we now have to compute various
gauge threshold corrections, which can be done quite analogously
to the computation of N=2 sector gauge thresholds described in
\cite{Lust:2003ky,Berg:2004ek,Akerblom:2007np,Blumenhagen:2007ip}.

First we have to compute the one-loop fluctuations around the $E1$ instantons, 
which are related to the threshold corrections
$\annstb{D5_r}{D9}$ and $\annstc{D5_r}{O9}$.
In the amplitude $\annstb{D5_r}{D9}$ for, say, the $D5$ brane
shown in figure \ref{poseeins} only the $(1+\Theta)/2$ insertions
in the trace give a non-vanishing contribution, as
$\Theta'$ and $\Theta''$ act by shifting the $D5$ along
the first two $T^2$s.
Denoting the Wilson lines along $D5$ as $(\alpha,\beta)$ and
transforming the expression into tree-channel we get the
general threshold corrections for this ${N}=2$ sector
\footnote{Note that in the following expressions we have already subtracted
the divergence due to tadpoles. In the full instanton amplitudes
the divergences cancel when taking all annulus and M\"obius diagrams into account.}
\bea
\label{intis}
    \anntb{D5}{D9}\, =\, \Re \left[ \annbb{E1}{D9} \right] \, &=&
    4\, \mathcal{T}_2\int_0^\infty  dl\  \sum_{{r,s=-\infty\atop (r,s)\ne (0,0)}}^{\infty}
         e^{-{\pi\, \mathcal{T}_2\, l\over \mathcal{U}_2}
         \left\vert r+\mathcal{U}s\right\vert^2 }\, e^{2\pi i (\alpha\, r+\beta\, s)}+\\
     &&+4\, \mathcal{T}_2\int_0^\infty  dl\ \sum_{r,s=-\infty}^\infty
         e^{ -{\pi\, \mathcal{T}_2\, l\over \mathcal{U}_2}
         \left\vert r+{1\over 2}+\mathcal{U}s\right\vert^2}\, 
          e^{ 2\pi i (\alpha\,(r+{1\over 2})
         +\beta\, s)} \; .\nonumber 
\eea
The appearing integrals are straightforward to compute using the
methods from \cite{Lust:2003ky,Akerblom:2007np,Blumenhagen:2007ip}.

The threshold corrections arising from the M\"obius strip
amplitude only receive contributions from the undressed
$\Omega$ insertion in the trace, as again $\Omega\Theta'$ and
$\Omega\Theta''$ shift the position of the $D5$-brane while
$\Omega\Theta$ is an ${N}=4$ sector and therefore
its threshold corrections  vanish.
It only remains to compute the integral 
\bea
     \anntc{D5}{O9}\, =\, \Re \left[ \annbc{E1}{O9} \right] \, &=&
   \mathcal{T}_2\int_0^\infty  dl\  \sum_{{r,s=-\infty\atop (r,s)\ne
        (0,0)}}^{\infty}  e^{-{\pi\, \mathcal{T}_2\, l\over \mathcal{U}_2}
         \vert r+\mathcal{U}s\vert^2 }\; .
 \label{schwachsinn}
\eea 
For an instanton $E1^{i,k}_{2,3,4}$ all these amplitudes do
not depend on $k$ and on $i$ only via the changed argument
$\mathcal{U}=\mathcal{U}_i$, while the functional form does  depend
on the discrete Wilson line.
For the three types of discrete Wilson lines we obtain:

\vspace{0.4cm}
\noindent
$\bullet\  E1_{2}$

\vspace{0.2cm}
\noindent
For the $E1$-$D9$ annulus diagram we get\footnotemark
\bea
     \annbb{E1_2}{D9} =-16\, \log\left( {\vartheta_2\over \eta}(2\mathcal{U}) \right)
\eea
where the prefactor of $16$ originates  from the 32 $D9$-branes
and  for the M\"obius strip we get the simple result\footnotemark
\bea
      \annbc{E1_2}{O9}=4\, \log\left( \eta (\mathcal{U}) \right)\; .
\eea

\addtocounter{footnote}{-1}
\footnotetext{Strictly speaking, \eqref{intis} and \eqref{schwachsinn} only yield the real parts of
the following expressions, which we believe to have its reason in the
fact that only the even spin structures have been included. As we are interested
in the gauge kinetic function, we promote the results to holomorphic functions.}
\addtocounter{footnote}{1}
\footnotetext{Note that in order to obtain this result, one has to remove
the massless open strings from \eqref{schwachsinn} (which is the correct
prescription for evaluating the amplitude \cite{Blumenhagen:2006xt}). In tree channel
this is done by regularising the integral in \eqref{schwachsinn} and removing the regulator
after subtracting the divergence.}
\vspace{0.4cm}
\noindent
$\bullet\  E1_{3}$

\vspace{0.2cm}
\noindent
For this instanton the computation of the $E1$-$D9$ annulus
is up to one  overall sign
in front of the second term in (\ref{intis}) identical to the former
case and we get 
\bea
    \annbb{E1_3}{D9}  =-16\, \log\left( {\vartheta_3\over \eta}(2\mathcal{U}) \right)\; .
\eea
For the M\"obius strip we get the same result as for the
 $E1_2$ instanton.

\vspace{0.4cm}
\noindent
$\bullet\  E1_{4}$

\vspace{0.2cm}
\noindent
Here  the annulus diagram yields
\bea
     \annbb{E1_4}{D9} =-16\, \log\left( {\vartheta_4\over \eta}(2\mathcal{U}) \right)
\eea
and again the M\"obius strip amplitude is not changed.

\vspace{0.4cm}
\noindent
Comparing this with the heterotic single instanton contribution in 
(\ref{singlea})-(\ref{singlec}), we realise that the sum there
corresponds to the three different classes of $E1$ instantons and that
the Type I one-loop determinants  \ 
$\exp\bigl[-\annsbb{E1}{D9}- \annsbc{E1}{O9}\bigr]$  \
precisely give the first two factors in each line.
It remains to compute the zero mode absorption
annulus diagrams $\annstf{D9}{E1}$ for the three types 
of $E1$-instantons.
Recall from section
2.2 that these amplitudes are related
to second derivatives of threshold corrections
${\partial^2\over \partial m^2} \annstb{D5}{D9}$.
We have just computed the latter.
Upon including a relative (continuous) Wilson line $m$ they
take the form
\bea
     \anntb{D5_r}{D9}=-16\, \log\left({\vartheta_r(m,2\mathcal{U})\over \eta(2\mathcal{U})}\right)
\eea
for $r=2,3,4$. 
Therefore, for the three different classes of single  $E1$ 
instantons, we get 
\bea
    && \int d^2\theta\, d^2 \mu\ \anntf{D9}{E1_2}\, \simeq\,  {\vartheta_2''\over \vartheta_2}(2\mathcal{U})
    =-{\pi^2\over 3}\left(E_2+\vartheta_3^4+\vartheta_4^4\right)(2\mathcal{U}) \nonumber\\
&& \int d^2 \theta \, d^2 \mu\ \anntf{D9}{E1_3}\, \simeq\,  {\vartheta_3''\over \vartheta_3}(2\mathcal{U})
     =-{\pi^2\over 3}\left(E_2+\vartheta_2^4-\vartheta_4^4\right)(2\mathcal{U}) \\
    && \int d^2 \theta\, d^2 \mu\ \anntf{D9}{E1_4}\, \simeq\,   {\vartheta_4''\over \vartheta_4}(2\mathcal{U})
     =-{\pi^2\over 3}\left(E_2-\vartheta_2^4-\vartheta_3^4\right)(2\mathcal{U})
     \nonumber \, ,
\eea
which agree
precisely with the third factor in each of the three
heterotic contributions in (\ref{singlea})-(\ref{singlec}).

We conclude that we managed to quantitatively match  the
holomorphic part of the single instanton contributions from the
heterotic and  the Type I side.
We consider this as evidence that the (single) D-brane instanton
calculus is correct and that in particular our  formulas (\ref{strangef},
\ref{strangeg})
relating the four fermionic zero mode absorption annulus diagrams
to second derivatives of gauge threshold corrections 
make sense.
Both on the heterotic and the Type I side there are non-holomorphic 
corrections in the string diagrams, arising as effects of
the massless modes.
It would be interesting to see whether the instanton
calculus on both sides gives complete agreement
also for the entire  string expressions and not
just for their holomorphic Wilsonian pieces.

From the Type I perspective it is clear that the three
contributions in (\ref{singlea})-(\ref{singlec})
originate  from three different $O(1)$ instantons distinguished 
by relative discrete Wilson lines.
However, on the heterotic side the S-dual interpretation
might  not be so familiar,
as the three terms are related to the GSO projection
of the $32$ left-moving world-sheet fermions. 
However, it was 
already  observed in \cite{Polchinski:1995df}
that the heterotic world-sheets with the four-different spin structures
of the $32$ free fermions are related to the four possible
$\mbb Z_2$ Wilson-lines for the $O(1)$ D1-instantons.
Indeed the fermions with PP-spin-structure do have
zero modes, so that they do not contribute
in  (\ref{singlea})-(\ref{singlec}) leaving only
the three contributions AA, AP, PA.

\subsection{Multiply wrapped single instantons }

Having identified the single instanton contributions,
it remains to clarify what the Type I S-dual of all the
higher terms in the heterotic instanton expansion is.
Very similar expressions have already appeared
for other heterotic instanton corrections 
(see \cite{Bachas:1997mc,Kiritsis:1997hf,Bianchi:2007rb}),
where it was pointed out that the higher terms
are S-dual to certain bound states of $E1$-instantons, which
can also be considered as single but  multiply wrapped $E1$-instantons \cite{Bachas:1997xn,Gava:1998sv}.

\begin{figure}[ht]
\centering
\hspace{10pt}
\includegraphics[width=0.7\textwidth]{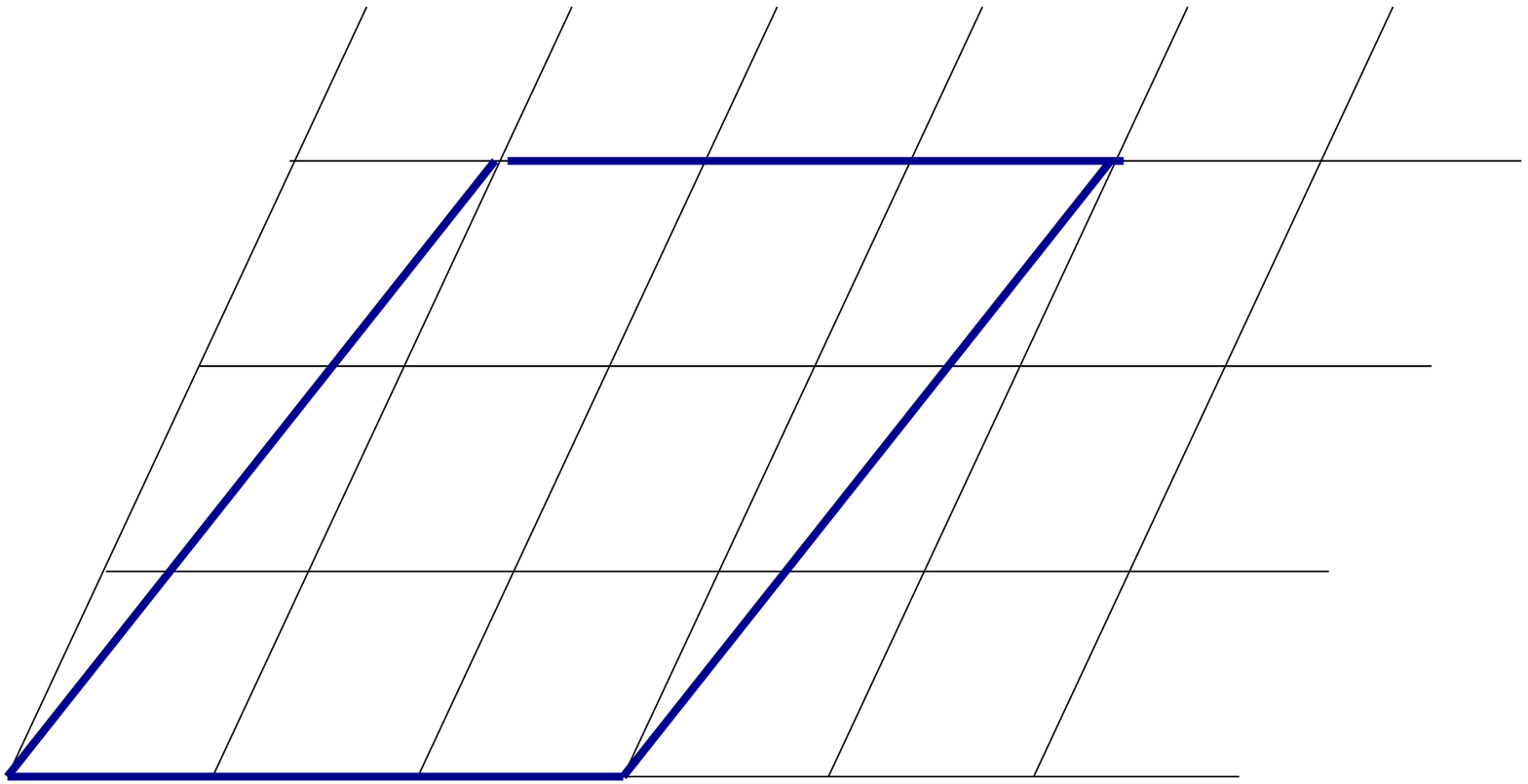}

\begin{picture}(100,1)
\put(-45,130){$p$}
\put(-53,5){$j$}
\put(25,5){$2k$}

\end{picture}

\vspace{-10pt}
\caption{Multiply wrapped $E1$ instanton for $k\in\mbb Z+{1\over 2}$
and $j,p\in\mbb Z$.\label{latta}}
\end{figure}

In order to determine the contributions of these multiply wrapped instantons to the
gauge kinetic function, one has to compute annulus and M\"obius diagrams.
They only differ from those appearing in the first order  instanton sector
in that the momentum/winding sums are changed. The necessary modification of these
sums is most easily described by introducing the effective complex structure modulus
$\mathcal{U}^{eff}$, which is just the complex structure modulus of the cycle the instanton wraps.
This cycle covers one of the three two-tori that constitute the internal manifold several times.
In other words, the lattice that defines the two-torus of the compactification manifold is
a sublattice of the lattice that describes the cycle the instanton wraps. This is illustrated in figure
\ref{latta}.
The equation \eqref{heteroricKKWsum} suggests that the effective complex structure modulus of the
cycle wrapped by the instanton that captures the contribution corresponding to a particular
matrix $A$ in the heterotic picture is encoded in
\bea
 \left(\begin{matrix} 1 \, ,\, \mathcal{U} \end{matrix} \right) A =
 \left(\begin{matrix} A_{11} + A_{21} \mathcal{U} \, ,\, A_{12} + A_{22} \mathcal{U} \end{matrix} \right) ,
\eea
where $\mathcal{U}$ is the complex structure modulus of the torus the instanton wraps several times.
More precisely, $\mathcal{U}^{eff}=(A_{12} + A_{22} \mathcal{U})/2(A_{11} + A_{21} \mathcal{U})$. 
Furthermore, it turns out that, in order to get a matching of the heterotic and type I results, one
has to modularly transform the matrix $A$ such that $A_{11}\in\mathbb{Z}+1/2$ and $A_{12}\in\mathbb{Z}$.

The matrices of interest for the present case are given in \eqref{matriceskjp} and we will now
discuss the three cases distinguished by whether $k$ and $j$ are integer or half-integer.
If $k\in\mathbb{Z}+1/2$ and $j\in\mathbb{Z}$ the matrix already has the correct structure and
the effective complex structure modulus is given by $\mathcal{U}^{eff}=(j+p\mathcal{U})/2k$. The cycle the instantons
wraps in this case is shown in figure \ref{latta}. The annulus and M\"obius diagrams
are those that appear in the one-instanton sector with $\mathcal{U}$ replaced by $\mathcal{U}^{eff}$. One therefore finally finds
that the instanton yields the contribution \eqref{hetotwo} to the heterotic amplitude.

If both $k$ and $j$ are half integers, one performs a modular T$^{-1}$ transformation on $A$ and obtains
$\mathcal{U}^{eff}=(j-k+p \mathcal{U})/2k$. As in the previous case, the annulus and M\"obius diagrams are those of
the one-instanton amplitudes, so one finds the one-instanton result with $\mathcal{U}$ replaced by $\mathcal{U}^{eff}=(j-k+p \mathcal{U})/k$,
i.e. \eqref{hetotwo} with argument$(j-k+p \mathcal{U})/k$. Applying a modular T transformation this can be
rewritten and reproduces the term \eqref{hetothree} in the heterotic amplitude.

Finally, if $k\in\mathbb{Z}$ and $j\in\mathbb{Z}+1/2$ one finds, after a modular S transformation on $A$,
$\mathcal{U}^{eff}=-k/2(j+p\mathcal{U})$. Another modular S transformation, this one on the full instanton amplitude, gives
the required result, the third term \eqref{hetoone} of the heterotic amplitude.

\subsection{The two-instanton sector}

\begin{figure}[ht]
\begin{center}
\hspace{-1cm}
\hbox{
\includegraphics[width=0.4\textwidth]{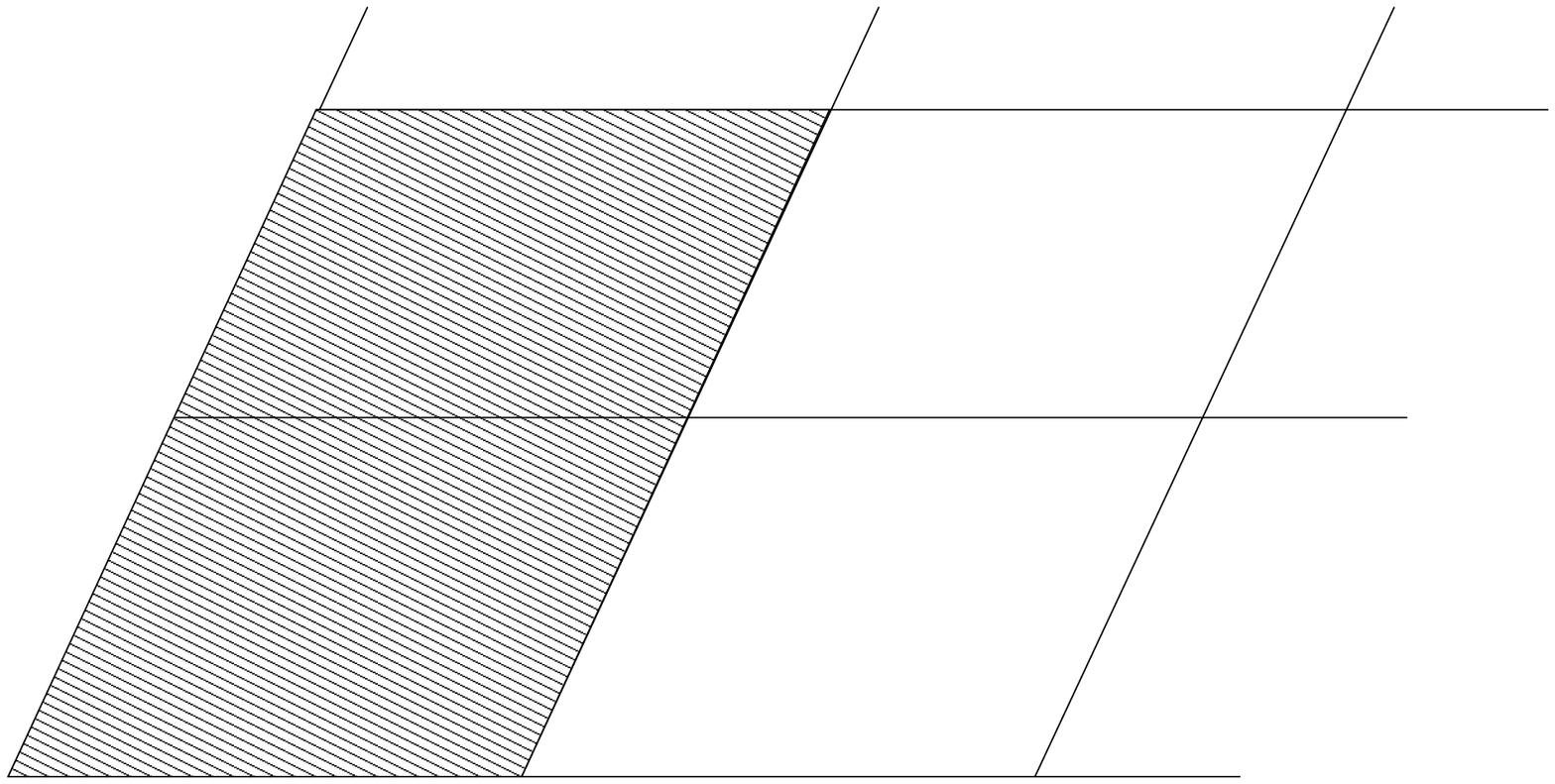}
\begin{picture}(100,1)
\put(-178,-6){$0$}
\end{picture}

\hspace{-2cm}
\includegraphics[width=0.4\textwidth]{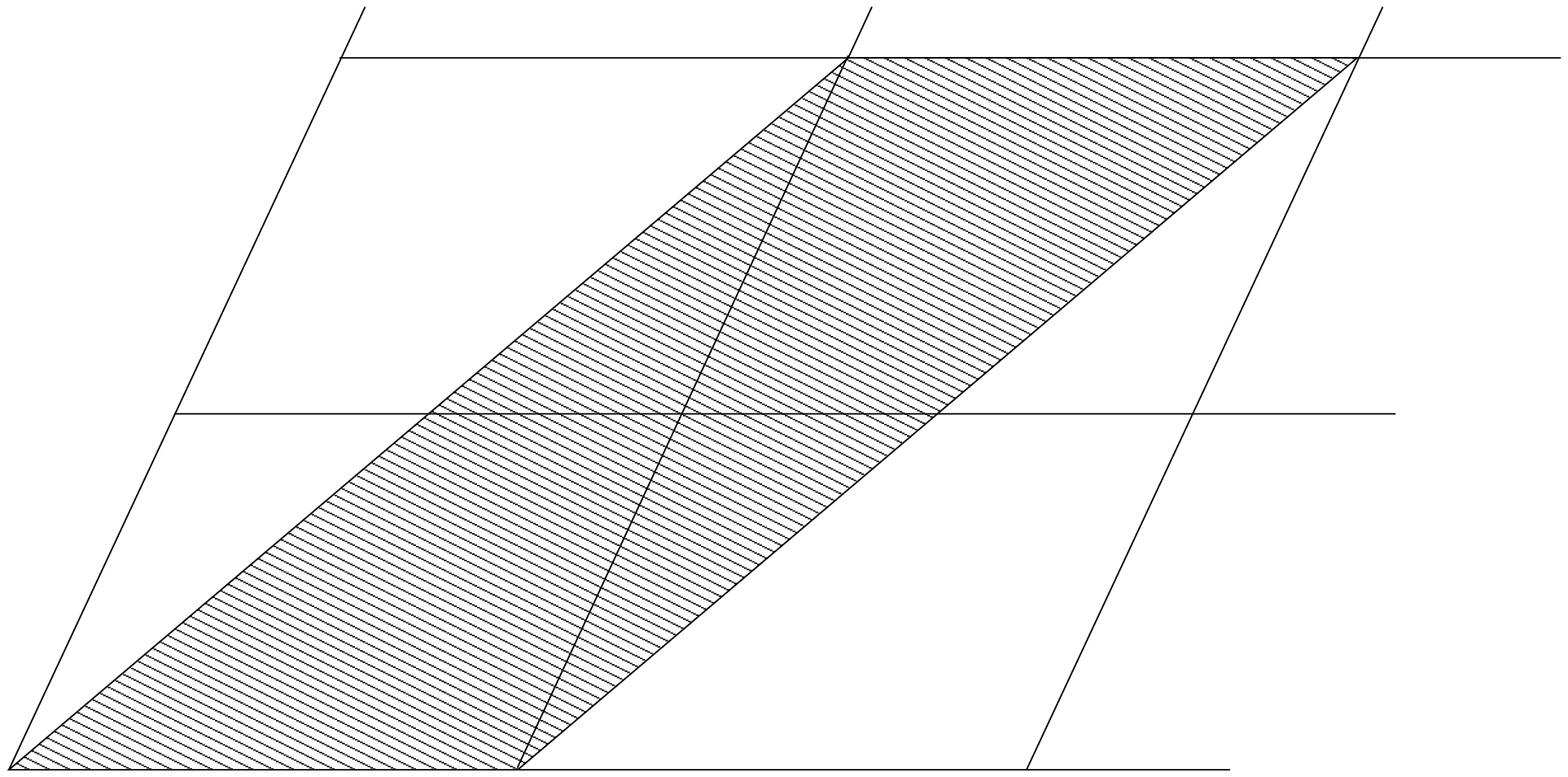}
\begin{picture}(100,1)
\put(-115,-10){$0$}
\end{picture}
}
\end{center}
\vspace{-10pt}
\caption{Left: $E1$ instanton for 
$k={1\over 2}$, $j=0$, $p=2$ . Right: $E1$ instanton for 
$k={1}$, $j={1\over 2}$, $p=1$
\label{lattbc}}
\end{figure}

Let us now look more closely at the second order  instanton sector.
The two heterotic orbits characterised by
\bea
         A=\left(\begin{matrix} {1\over 2} & 0 \\ 0 & 2
            \end{matrix} \right),\quad\quad
          A=\left(\begin{matrix} 1& {1\over 2}  \\ 0 & 1
            \end{matrix} \right)\;          
\eea
correspond to the twice-wrapped $E1$-instantons shown in figure
\ref{lattbc}.

\noindent
However, what about the various Type I poly two-instanton contributions
\bea
 E1^{i,k}_{3}-E1^{i',k'}_{4},\quad
      \quad  
     E1^{i,k}_{2}-E1^{i',k'}_{4}\quad
      \quad  
      E1^{i,k}_{2}-E1^{i',k'}_{3}\; 
\eea
which arise at the same second order in $\exp(2\pi i\mathcal{T})$ and
which in section 2 we proposed to exist? 

First, following our $E1$-instanton calculus introduced in section 2,
let us compute
these genuine two instanton contributions explicitly. We note
that for $i\ne i'$ the zero mode absorption amplitude $\annsbf{i}{i'}$
vanishes, as the two instantons wrap different $T^2$s so that
the threshold correction cannot depend on any of the Wilson-lines
along the cycles of the two instantons. Similarly, for
$i=i'$ but $k\ne k'$ the amplitude $\annsbf{k}{k'}$ vanishes
because there are no momentum modes invariant under
the $\Theta$ insertion. Therefore, we only have to consider
the case $i=i'$ and $k=k'$.
For instance, the first amplitude (up to a normalisation factor $\kappa$)
we expect to be equal to
\bea
\label{steiger}
 \Lambda\left(E1_{3},E1_{4}\right)&=& \kappa\,
  \int d^4 x_{rs}\, d^2\theta_r \, d^2\theta_s\, d^2\mu_r\, d^2\mu_s \ 
  \left(  \anntf{D9}{E1_3}\, \annbf{E1_3}{E1_4}+
        \anntf{D9}{E1_4}\, \annbf{E1_4}{E1_3} \right) \nonumber \\
 && \hspace{-1.2cm} \exp\left(-\annbb{E1_3}{D9} - \annbc{E1_3}{O9}\! \right) \,
   \exp\left(-\annbb{E1_4}{D9} - \annbc{E1_4}{O9}\! \right)\ \, e^{2\pi i\, \mathcal{T}}
\eea
In order to determine $\annsbf{E1_3\,}{\, E1_4}\hspace{3pt}=\hspace{3pt}\annsbf{E1_4\,}{\, E1_3}\,$,
we only have to compute the
second derivative of $\annsbt{D5_3\,\,}{\,\, D5_4}$.
As the $E1_3$ and $E1_4$ instantonic branes
are parallel on all three
$T^2$s,  only the
term with the $\Theta$ insertion (in loop channel) contributes 
to the threshold correction amplitude $\annsbt{D5_3\,\,}{\,\,D5_4}$, and we find
\bea
 \label{googlebombe}
   \annbt{D5_3}{D5_4}=\log\left( {\vartheta_2\over
        \vartheta_3}(2\mathcal{U})\right)
\eea
leading to
\bea
   \int d^4 x_{rs}\,\, d^2\theta_s\, d^2\mu_s \ \annbf{E1_3}{E1_4}
 \simeq {\partial^2\over \partial m^2}
       \annbt{D5_3}{D5_4}\biggr\vert_{m=0} 
   \simeq {\vartheta''_2\over \vartheta_2} - {\vartheta''_3\over
     \vartheta_3}     = -\pi^2\,
        \vartheta_4^4(2\mathcal{U})\; .
\eea
Collecting all terms, we find for  this two instanton amplitude
\bea
\label{schalke}
     \Lambda\left(E1_{3},E1_{4}\right)={\pi^4 \kappa\over 3} \,
    e^{2\pi i \mathcal{T}}\ {\vartheta_4^4(2\mathcal{U})\over \eta^8(\mathcal{U}) }
      \left({\vartheta_3\vartheta_4\over \eta^2}(2\mathcal{U})\right)^{\!\! 16}
     \! (2\, E_2 -\vartheta^4_3-\vartheta^4_4)(2\mathcal{U})\; , \quad
\eea
which invoking some $\vartheta$-function identities can be written as
\bea 
  \Lambda\left(E1_{3},E1_{4}\right)={4\pi^4\kappa \over 3}  \, 
    {e^{2\pi i\, \mathcal{T}}\over  \eta^4(2\mathcal{U})} \, \, 
   \left( {\vartheta_4 \over \eta}(4\mathcal{U})\right)^{\!\! 16}
   \left( E_2-\vartheta_2^4-\vartheta_3^4\right)(4\mathcal{U})\; .
\eea
One realises that this has precisely the form of the holomorphic part of the heterotic contribution
(\ref{wichta}), which naively could mean that this poly-instanton
contribution is already included in the heterotic expression
of multiply wrapped single instantons.

The other possibility is that the equality is more a coincidence
in the sense  that a
D-brane wrapped twice around a cycle $b$ has (up a to normalisation
factor two) the same partition function
as a pair of singly wrapped D-branes with relative Wilson
line $w={1\over 2}$ along the cycle $b$.
This would explain why the two-instanton sector
$E1_{3}-E1_{4}$  has the same functional form as the
doubly wrapped one instanton $E1_{4}$.
In this latter case, the fact that we really get the same
result for the two-instanton and single doubly-wrapped instanton
corrections establishes another positive test of the formulas
\eqref{steiger} and \eqref{googlebombe}.

Before we draw our final conclusions, let us proceed and
collect more data. 
Completely analogously, using
\bea
   \int d^4 x_{rs}\,\, d^2\theta_s\, d^2\mu_s\  \annbf{E1_2}{E1_4}
     \simeq \pi^2\, \vartheta_4^4(2\mathcal{U})\; 
\eea
we can compute the $E1_2-E1_4$ poly-instanton
correction and obtain 
\bea
\label{bochum}
    \Lambda\left(E1_{2},E1_{4}\right)=-{\pi^4\kappa \over 3}  \, 
    e^{2\pi i \mathcal{T}}\ {\vartheta_4^4(2\mathcal{U})\over \eta^8(\mathcal{U}) }
      \left({\vartheta_2\vartheta_4\over \eta^2}(2\mathcal{U})\right)^{\!\! 16}
 \!   (2\, E_2 -\vartheta^4_2+\vartheta^4_4)(2\mathcal{U}) \quad
\eea
which can be expressed as
\bea
   \Lambda\left(E1_{2},E1_{4}\right)= {\pi^4\kappa \over 3} \, e^{- \pi i /3}
   { e^{2\pi i\, \mathcal{T}}\over  \eta^4(2\mathcal{U})} \, \, 
   \left( {\vartheta_2 \over \eta}({\textstyle{1\over 2}}+\mathcal{U})\right)^{\!\! 16}
   \left( E_2+\vartheta_3^4+\vartheta_4^4\right)({\textstyle{1\over 2}}+\mathcal{U})\; .
 \nonumber
\eea
This expression is equal to the heterotic contribution
(\ref{wichb}) arising from  the other orbit.  
These two singly wrapped instantons
have the same partition function as the doubly wrapped single instanton
$E1_{2}$. Note, that with the modular properties (\ref{ufos})
it is obvious that under $U^2$ the two contributions
(\ref{bochum}) and (\ref{schalke}) get exchanged.

It remains to discuss the  $E1_2-E1_3$ poly-instanton
correction. However, since these two instantons are only different
by a relative Wilson line $(\alpha,\beta)=(1,0)$, there appear
extra zero modes that are not all removed
by the $\Theta$ projection. This is immediately obvious
by looking at the related $\annstb{D5_2\,\,}{\,\,D5_3}$ amplitude
which tells us that there are four fermionic charged matter zero modes
in the $D5_2$-$E1_3$ sector, so that
there is no $E1_3$ instanton correction to the gauge
coupling on the $D5_2$ brane.
By noting that $(\vartheta_2\vartheta_3/\eta^2)(2\mathcal{U})=(\vartheta_2/\eta)(\mathcal{U})$,
we expect that, if this coupling were there, it would be related
to the heterotic contribution from the string doubly wrapped
along the $x$-axis. However, this contribution is
absent on the heterotic side, which again is consistent
with the absence we just observed on the Type I side.

In the second order  instanton sector it was possible to consistently relate
all poly-instanton contributions to heterotic contributions
from twice wrapped single instantons. 
When finally briefly discussing the third order  instanton sector,
we will see that this agreement was merely
a coincidence, due to the equality (up to a normalisation factor
two)  of the partition
functions of a doubly wrapped
brane and a pair of  singly wrapped branes with relative one-half Wilson lines.

Note that by S-duality these two-instanton contributions
are expected to arise from two world-sheet instantons with 
different spin structures of the $SO(32)$ fermions.

\subsection{The three-instanton sector}
To clarify the relation  between multiply wrapped single instantons
and poly-instantons let us look more closely at the third order
instanton sector.
The heterotic gauge kinetic function receives the four contributions
\bea
\label{threehetinst}
    \Lambda_3 (\mathcal{U},\mathcal{T})&=&{2\over 3}\,{\cal A}\left[\begin{matrix} 0\\ 1 \end{matrix} \right]
\left( {6\mathcal{U}} \right)\, \, e^{3\pi i\,\mathcal{T}} +
   {2\over 3}\,{\cal A}\left[\begin{matrix} 0\\ 1 \end{matrix} \right]
\left( {2\mathcal{U}\over 3} \right)\, \, e^{3\pi i\, \mathcal{T}}+ \\
&&
{2\over 3}\,{\cal A}\left[\begin{matrix} 0\\ 1 \end{matrix} \right]
\left( {2+2\mathcal{U}\over 3} \right)\, \, e^{3\pi i\,
  \mathcal{T}}+
{2\over 3}\,{\cal A}\left[\begin{matrix} 1\\ 1 \end{matrix} \right]
\left( {1+2\mathcal{U}\over 3} \right)\, \, e^{3\pi i\,
  \mathcal{T}} \; . \nonumber
\eea
Looking for instance at the $\vartheta$-functions
at arguments which are multiples or quotients of $2\mathcal{U}$ by three,
one realises that these triply  wrapped instantons
can actually   only be  equivalent to the
product of three singly wrapped   branes
with discrete Wilson lines quantised in units of ${1\over 3}$ but
but not to poly 3-instanton sectors with relative discrete Wilson lines
$w\in\{0, 1/2,1 \}$.

On the Type I side we expect to get all kinds
of instanton corrections. First, there will be
the triply wrapped single instanton contributions
directly present on the heterotic side (\ref{threehetinst}).
Moreover, there are poly-instanton contributions involving
one doubly wrapped instanton and a second singly wrapped instanton.
In addition, there are poly-instanton contributions from
three single instantons, which are partly
already a consequence of the power tower like behaviour
starting at the two-instanton level, i.e. the third
order ($n=2$) terms in the expansion (\ref{multiinst}).

One genuine three instanton sector, which is first present
at this level involves all three types of single instantons
$E_{2,3,4}$, all wrapping the same $T^2$ and localised
on the same transversal  pairs of fixed points.

We will now compute this amplitude using the methods developed in section 2.
After summing over all
non-vanishing combinations for absorbing the 
zero modes it reads
\bea
\label{steigerdrei}
 \Lambda\left(E1_2,E1_{3},E1_{4}\right)&=& \kappa\,
  \int d^4 x_{rs}\,\, d^4 x_{su}\, \, d^2\theta_r \,
  d^2\theta_s\,d^2\theta_u\, d^2\mu_r\, d^2\mu_s \,d^2\mu_u\ \,\, e^{3\pi i\,
  \mathcal{T}}  \nonumber\\
  &&  \hspace{-3.4cm} 
\left(  \anntf{D9}{E1_2}\ \annbf{E1_2}{E1_4}\ \annbf{E1_4}{E1_3}
 +  \anntf{D9}{E1_3}\ \annbf{E1_3}{E1_4}\ \annbf{E1_4}{E1_2} +
 \anntf{D9}{E1_4}\ \annbf{E1_4}{E1_2}\ \annbf{E1_4}{E1_3}
 \right)\  \nonumber \\
&&\hspace{-3.7cm}  \exp\left(-\annbb{E1_2}{D9} - \annbc{E1_2}{O9}\! \right) 
  \  \exp\left(-\annbb{E1_3}{D9} - \annbc{E1_3}{O9}\! \right)\ 
  \exp\left(-\annbb{E1_4}{D9} - \annbc{E1_4}{O9}\! \right)  \nonumber
\eea
We have argued before that, due to charged zero modes, there are no $E1_2$($E1_3$) instanton corrections
to the gauge coupling on $D5_3$($D5_2$). We therefore do not expect $E1_2$ and $E1_3$ to mutually correct
their instanton actions. That is why the diagrams $\annsbf{E1_2\,}{\,E1_3}$ and $\annsbf{E1_3\,}{\,E1_2}$ are not allowed
in the above expression, which is reflected in a divergence that shows up when calculating these diagrams with the
methods used before.

All  the ingredients of the amplitude shown above have already been computed, so we
proceed by simply inserting them and find
\bea
\label{steigerdreia}
 \Lambda\left(E1_2,E1_{3},E1_{4}\right)&=& - \kappa\, \pi^4
  \, {\vartheta_4^8(2\mathcal{U})\over \eta^{12} (\mathcal{U})} \left(
  { \vartheta_2\,  \vartheta_3\,  \vartheta_4 \over \eta^3}(2\mathcal{U})
  \right)^{\!\! 16}\, \sum_{r=2,3,4} {\vartheta_r''\over \vartheta_r} (2\mathcal{U})
   \  e^{3\pi i\, \mathcal{T}}         \nonumber \\
  &=& \kappa\, {\pi^6\, 2^{18} }
  \, { E_2\over \vartheta_2^2\, \vartheta_3^2} (2\mathcal{U})\  e^{3\pi i\,
  \mathcal{T}}  \, ,
\eea
where we have used the identity $2\eta^3=\vartheta_2\,  \vartheta_3\,
  \vartheta_4$.  Note that this poly-instanton contribution
is invariant under the modular group $\Gamma_2^{\mathcal U}$.
This amplitude eventually has  a comparably  simple
functional form and is not present on the heterotic
side. Therefore, in contrast to the two poly-instanton
contributions this three poly-instanton amplitude
can apparently not be considered equivalent to
a triply wrapped single
instanton but is genuinely new.

We have argued that the instantons $E1_2$ and $E1_4$ mutually correct their actions
and that the same is true for the instantons $E1_3$ and $E1_4$. We have also explained
that this should not be the case for $E1_2$ and $E1_3$. Taking this into account,
the infinite power tower \eqref{pentenried} of instanton corrections
to the gauge kinetic function in the present case takes
the following form (shown explicitly up to third order):
\bea
 \delta f &=& \anntf{D9}{E1_2}\ \exp 
       \left( - S_2 + \annbf{E1_2}{E1_4} \ 
       e^{-S_4\ +\hspace{5pt}\annsbf{E1_4\,}{\,E1_2} \  \ {e^{-S_2}}^{\Ddots}\
               +\hspace{5pt}\annsbf{E1_4\,}{\, E1_3} \ \  {e^{-S_3}}^{\Ddots} } \right)
 \\ &&\hspace{-0.3cm}  + \anntf{D9}{E1_3}\ \exp 
       \left( - S_3 + \annbf{E1_3}{E1_4} \ 
       e^{-S_4 \ +\hspace{5pt}\annsbf{E1_4\,}{\, E1_3} \ \  {e^{-S_3}}^{\Ddots}
               \ +\hspace{5pt}\annsbf{E1_4\,}{\, E1_2} \ \  {e^{-S_2}}^{\Ddots} } \right)
 \nonumber \\ &&\hspace{-0.9cm} + \anntf{D9}{E1_4}\ \exp 
       \left( - S_4 \ + \annbf{E1_4}{E1_3}\ 
         e^{-S_3\  +\hspace{5pt}\annsbf{E1_3\,}{\, E1_4} \ \ {e^{-S_4}}^{\Ddots} } 
                    + \annbf{E1_4}{E1_2}\ e^{-S_2\ +\hspace{5pt}\annsbf{E1_2\,}{\, E1_4} \ \  {e^{-S_4}}^{\Ddots}} \right) \nonumber
\eea
where we have not shown explicitly the one-loop determinants. 
When expanding this power tower one finds that the first and second order terms reproduce the one- and two-instanton
amplitudes we wrote down before. Furthermore, the third order terms involving all three different instantons
equal the three-instanton amplitude we just computed, including the correct relative combinatorial prefactors
of the three different terms.

Note that in the complete power tower expression for the gauge kinetic function
also the multiply wrapped instantons have to be included. We expect that this eventually gives
the full exact non-perturbative result.
The expectation is that this whole (sort of fractal)
object is modular invariant under $\Gamma_2^{\mathcal{U}}$. It is beyond the scope
of this paper to further elucidate the mathematical aspects, like
modular and convergence  issues, 
of these power towers.

Moreover, after integrating out all massive modes, the pure 
$N=1$  $SO(32)$  Yang-Mills theory on the D9-branes
is expected to show gaugino condensation and a dynamically
generated superpotential
\bea
\label{adssup}
         W=  A\   e^{ {2\pi i\, \mathcal{S}\over 30\,} }\; ,
\eea
where the one-loop beta-function coefficient is already included.
However, as we have just seen, in string theory the tree-level
gauge coupling $f_{\rm tree}=i\, \mathcal{S}$ receives further 
one-loop threshold and instanton corrections so that we expect the whole
power tower to appear in the exponent of \eqref{adssup}.
Therefore, the heterotic - Type I model also serves -- in some sense -- as an example
of the poly-instanton superpotential corrections discussed
in section \ref{secsuperpot}. 

Let us summarise our conclusions from the 
very explicit discussion of the instanton corrections to
the holomorphic gauge kinetic function of this 
heterotic-Type I dual orbifold model.

\begin{itemize}
\item The ordinary gauge threshold computation for the gauge coupling
    of an ${N}=1$ supersymmetric heterotic string model
    includes only instanton corrections from multiply wrapped
    though single world-sheet instantons distinguished in our model by the
    spin structures of the $32$ left-moving fermions. 

\item On the  Type I dual side these corrections arise from
     multiply wrapped single $E1$-instantons with different
      $\mbb Z_2\subset U(1)$ Wilson lines. Here, we do not 
    encounter any obstruction to the computation of  
      poly-instanton  corrections. In fact the 
   relevant  zero mode absorption amplitudes could be
    computed explicitly and for $E1_2$-$E1_4$ and $E1_3$-$E1_4$ 
        were  non-vanishing.

\item For the special cases that the instantons in a poly-instanton amplitude can be considered (in the aforementioned sense)
       as a single multiply wrapped instanton, 
    the two resulting  instanton amplitudes agree, which gives 
     us some degree of confidence that the euclidean instanton
    calculus presented in section 2 is correct.

  \item In view of these results and assuming that S-duality holds,
       two logical possibilities 
        seem to offer themselves. Either on the Type I side, we are missing a 
     further criterion for instantons to contribute or
   the heterotic computation involving a sum over 
     oscillator, Kaluza-Klein and winding  excitations of a single heterotic
    string running in a loop     
    is blind against these poly-instanton contributions,
   as its starting point is per se a single heterotic string world-sheet.
   The fact that one is only dealing with one world-sheet is clear from
   \eqref{pelle}, as it instructs one to perform a trace
   in the Hilbert space of {\it one} CFT.
\end{itemize}

\vspace{0.6cm}
\section{Remarks}

We would like to close with a number of more general concluding remarks.

What we have exemplified and explained mostly for the holomorphic
gauge kinetic function, is expected to occur much more generally.
The following statement summarises what we have observed
in this paper: Whenever for certain couplings one finds
instanton corrections to instanton actions, one should get a 
power tower like proliferation of instanton corrections. 
Compared to world-sheet instantons in the heterotic string,
the existence of these poly-instanton corrections  is much more
evident for D-brane instantons, simply for the reason that here 
we can use open string theory to compute the
zero mode absorption diagrams involving many D-brane world-sheets,
in other words terms in the effective action of multiple D-branes.
Assuming S-duality,  for fundamental string instantons, 
these terms would presumably be
visible in  an approach allowing the treatment
of multiple disconnected string world-sheets and 
their higher order interactions. In analogy to $E1$-instantons,
these interactions
are not expected to be splitting and joining processes of strings, but rather terms
in the two-dimensional effective action of multiple string world-sheets.

It is important to emphasise that none of the poly-instanton 
corrections we proposed violates
the non-renorma\-li\-sation theorems for holomorphic couplings in
$N=1$ supersymmetric four-dimensional string compactifications, which
were originally  derived 
for instance in \cite{Dine:1986zy,Dine:1987bq}. 
In fact they generalise them to poly-string instantons.
It would be interesting to see, whether the vanishing instanton sums
of \cite{Beasley:2003fx} can be generalised to cases where
poly-instanton contributions do exist.

Clearly, it is important to know which instanton actions receive
 instanton corrections. We established this behaviour for $O(1)$ instantons
in $N=1$ supersymmetric orientifold models. 
By just looking at the zero mode counting, we do not expect
similar corrections to the 1/2-BPS fundamental, $E1$ and $E3$ instanton
corrections to the $N=2$ hypermultiplet moduli space as recently
discussed for instance in \cite{RoblesLlana:2006is, RoblesLlana:2007ae}. Similarly, these corrections are expected to be
absent for the topological A-model and its various genus $g$
world-sheet instanton corrections.

These poly-instanton corrections will also occur for
charged matter couplings in the superpotential. If
there are at least two instantons which correct the action of the rigid charged
instanton and which do not carry any charged matter 
zero modes, one gets the exponential proliferation we have seen
in this paper.
If however an instanton that corrects the action of an instanton
contributing to the superpotential
carries  additional  charged matter zero modes,
then for instance the  two instanton sector can contribute to a
different charged matter coupling constituting possibly the leading order term.

Even though these poly-instanton corrections are strongly
suppressed in the perturbative regime, they might, under certain circumstances,
provide  the leading order dependence on some
K\"ahler moduli. It remains to be seen what the effects of  such corrections
are for moduli stabilisation and fine-tuning problems.

\vskip 1cm 
 {\noindent  {\Large \bf Acknowledgements}} 
 \vskip 0.5cm 
We would like to thank Nikolas Akerblom, Dieter L\"ust, Sebastian Moster,
Erik Plauschinn, Timo Weigand and in particular
Emilian Dudas for discussions. 
This work is supported in part by the European Community's Human 
Potential Programme under contract MRTN-CT-2004-005104 `Constituents, 
fundamental forces and symmetries of the universe'. 
R.B. thanks the Erwin Schr\"odinger Institute for Mathematical Physics in Vienna
for hospitality.
%%%%%%%%%%%%%%%%%%%%%%%%%%%%%%%%%%%%%%%%%%%%%%% 
%%%%%%%%%%%%%%%%%%%%%%%%%%%%%%%%%%%%%%%%%%%%%%% 
%%%%%%%%%%%%%%%%%%%%%%%%%%%%%%%%%%%%%%%%%%%%%%% 
%%%%%%%%%%%%%%%%%%%%%%%%%%%%%%%%%%%%%%%%%%%%%%% 
%%%%%%%%%%%%%%%%%%%%%%%%%%%%%%%%%%%%%%%%%%%%%%% 
%%%%%%%%%%%%%%%%%%%%%%%%%%%%%%%%%%%%%%%%%%%%%%% 
%%%%%%%%%%%%%%%%%%%%%%%%%%%%%%%%%%%%%%%%%%%%%%% 
%%%%%%%%%%%%%%%%%%%%%%%%%%%%%%%%%%%%%%%%%%%%%%% 

\clearpage 
\nocite{*} 
\bibliography{rev} 
\bibliographystyle{utphys} 
%\bibliographystyle{plain} 

%%%%%%%%%%%%%%%%%%%%%%%%%%%%%%%%%%%%%%%%%%%%%%% 
%%%%%%%%%%%%%%%%%%%%%%%%%%%%%%%%%%%%%%%%%%%%%%% 
%%%%%%%%%%%%%%%%%%%%%%%%%%%%%%%%%%%%%%%%%%%%%%% 
%%%%%%%%%%%%%%%%%%%%%%%%%%%%%%%%%%%%%%%%%%%%%%% 
%%%%%%%%%%%%%%%%%%%%%%%%%%%%%%%%%%%%%%%%%%%%%%% 
%%%%%%%%%%%%%%%%%%%%%%%%%%%%%%%%%%%%%%%%%%%%%%% 
%%%%%%%%%%%%%%%%%%%%%%%%%%%%%%%%%%%%%%%%%%%%%%% 
%%%%%%%%%%%%%%%%%%%%%%%%%%%%%%%%%%%%%%%%%%%%%%% 

\end{document}